\documentclass[apjl, iop]{emulateapj}

\slugcomment{Accepted for publication in AJ}

\usepackage{color}

\shorttitle{Observation of SN 2012fr}
\shortauthors{Ju-Jia Zhang et al.}

\begin{document}
\title{Optical and Ultraviolet Observations of the \\ Narrow-Lined Type Ia SN 2012fr in NGC 1365 }

\author{Ju-Jia Zhang\altaffilmark{1,2,3,4},Xiao-Feng Wang\altaffilmark{2},Jin-Ming Bai\altaffilmark{1,3},Tian-Meng Zhang\altaffilmark{5,6}, Bo Wang\altaffilmark{1,3}, Zheng-Wei Liu\altaffilmark{1,3,7},Xu-Lin Zhao\altaffilmark{2},Jun-Cheng Chen\altaffilmark{2}}

\altaffiltext{1}{Yunnan Observatories (YNAO), Chinese Academy of Sciences, Kunming 650011, China.(jujia@ynao.ac.cn)}
\altaffiltext{2}{Physics Department and Tsinghua Center for Astrophysics (THCA), Tsinghua University, Beijing 100084, China.(wang\_xf@mail.tsinghua.edu.cn)}
\altaffiltext{3}{Key Laboratory for the Structure and Evolution of Celestial Objects, Chinese Academy of Sciences, Kunming 650011, China.(baijinming@ynao.ac.cn)}
\altaffiltext{4}{University of Chinese Academy of Sciences, Chinese Academy of Sciences, Beijing 100049, China.}
\altaffiltext{5}{National Astronomical Observatories of China (NAOC), Chinese Academy of Sciences, Beijing 100012, China.}
\altaffiltext{6}{Key Laboratory of Optical Astronomy, National Astronomical Observatories, Chinese Academy of Sciences, Beijing 100012, China.}
\altaffiltext{7}{Argelander-Institut für Astronomie, Auf dem Hügel 71, D-53121, Bonn, Germany.}
\begin{abstract}
Extensive optical and ultraviolet (UV) observations of the type Ia supernova (SN Ia) 2012fr are presented in this paper. It has a relatively high luminosity, with an absolute $B$-band peak magnitude of about $-19.5$ mag and a smaller post-maximum decline rate than normal SNe Ia [e.g., $\Delta m _{15}$($B$) $= 0.85 \pm 0.05$ mag]. Based on the UV and optical light curves, we derived that a $^{56}$Ni mass of about 0.88 M$_{\sun}$ was synthesized in the explosion. The earlier spectra are characterized by noticeable high-velocity features of \ion{Si}{2} $\lambda$6355 and \ion{Ca}{2} with velocities in the range of $\sim22,000$--$25,000$ km s$^{-1}$. At around the maximum light, these spectral features are dominated by the photospheric components which are noticeably narrower than normal SNe Ia. The post-maximum velocity of the photosphere remains almost constant at $\sim$12,000 km s$^{-1}$ for about one month, reminiscent of the behavior of some luminous SNe Ia like SN 1991T. We propose that SN 2012fr may represent a subset of the SN 1991T-like SNe Ia viewed in a direction with a clumpy or shell-like structure of ejecta, in terms of a significant level of polarization reported in \citet{Maund12frp}.
\end{abstract}


\section{Introduction}
\label{sect:Intro}
Type Ia supernovae (SNe Ia) are widely accepted as the results of thermonuclear explosion of accreting carbon-oxygen white dwarf (WD) with a mass close
to the Chandrasekhar limit ($\sim$1.4 M$_{\odot}$) in a binary system \citep{HiNi,Bow12,Maoz14}. They play important roles in many aspects of astrophysics, especially in observational cosmology because of their use as distance indicators probing the expansion history of the universe \citep{Riess98,Schmidt98,Perl99}. Observationally, most of the SNe Ia show strikingly similar photometric and spectroscopic behavior (i.e., \citealp{Sun96,Filip97}), and the remained scatter can be better understood in terms of an empirical relation between light-curve width and luminosity (i.e., \citealp{Phillip93}). Nevertheless, there is increasing evidence for the observed diversity that cannot be explained with such a relation. The representative subclasses include: (1) overluminous group like SN 1991T characterized by weak \ion{Si}{2} absorption and prominent iron in the near-maximum light spectra \citep{Filip92a,Phillip92}; (2) underluminous events like SN 1991bg that exhibit strong \ion{Si}{2} $\lambda$5972 and  $\sim$4000\AA~Ti features (i.e.,\citealp{Filip92b,Ben05}); (3) peculiar objects like SN 2002cx which have extremely low ejecta velocities and luminosities \citep{Li02cx}; (4) super-Chandrasekhar mass SNe Ia like SN 2007if \citep{Scalzo} and SN 2009dc\citep{Silver11}, which are characteristic of extremely high luminosity but relatively low expansion velocity. In addition, there are also reports of circumstellar interaction in some SNe Ia such as SN 2002ic \citep{Hamuy} and PTF 11kx \citep{Dilday}, though their classifications are still controversial because of bearing similarities to type IIn supernovae. The existence of above peculiar subtypes indicate that there are possibly multiple channels leading to SN~Ia explosions.

Besides peculiar SN Ia events, specific classification schemes have recently been proposed to highlight the diversity of relatively normal SNe Ia. For example, \citet{Ben05} found that normal SNe Ia can be further subclassified by the temporal velocity gradient of the \ion{Si}{2} $\lambda$6355 line, i.e., the group with a high-velocity gradient (HVG) and the group with a low-velocity gradient (LVG). Based on the equivalent width (EW) of the absorption features of \ion{Si}{2} $\lambda$5972 and \ion{Si}{2} $\lambda$6355, \citet{Branch06,Branch09} suggested dividing the SN Ia sample into four groups: cool (CL), shallow silicon (SS), core normal (CN), and broad line (BL); the CL and SS groups mainly consist of peculiar objects like SN 1991bg and SN 1991T, respectively. \citet{Wang09a} proposed using the expansion velocity of the \ion{Si}{2} $\lambda$ 6355 line to distinguish the subclass with a higher \ion{Si}{2} velocity (HV) from that with a normal velocity (NV). In spite of different criteria adopted in these classifications, the respective subsamples show some overlap with each other. For example, the FAINT subclass from \citet{Ben05}  tend to match the SN 1991bg/CL subclass. The HV subclass overlap with the BL and the HVG ones, as suggested by the Berkeley and CfA spectral datasets of supernova \citep{Blondin12,Silver12}. In particular, the HV SNe Ia are found to have redder $\bv$ colors  \citep{Wang09a} and different locations within host galaxies in comparison to the NV ones \citep{Wang13}, suggesting that the properties of their progenitors may be different.

SN 2012fr is a type Ia supernova discovered at a relatively young phase in nearby galaxy NGC 1365 \citep{Klotz,CEBT3277}. Owing to its brightness, extensive follow-up observations were performed in multi-wavebands for this object immediately after the discovery. Childress et al. (2013, hereafter C13) have presented observations of earlier optical spectra for SN 2012fr, showing clear signatures of high-velocity features (HVFs) that are detached from the photospheric components. After comparing with subclasses of SNe Ia defined in different classification schemes, they suggested that SN 2012fr may represent a transitional event between nominal spectroscopic subclasses of SN Ia, with important dissimilarities with the overluminous SN 1991T-like subclass of SNe Ia. \citet{Maund12frp} presented the spectropolarimetric observations of this object, spanning from $-$11 days to +24 days with respect to $B$-band maximum light. They found that the high-velocity components of the spectral features are highly polarized in the earlier phases but the polarization decreases as these features become weaker. However, the continuum polarization for the SN is always low $<$0.1\%, suggestive of an overall symmetry of the photosphere for SN 2012fr.

In this paper, we present extensive optical and ultraviolet (UV) observations of SN 2012fr. The large dataset of SN 2012fr can help us further understand diversity of SNe Ia, its physical origin, and impact on cosmological applications. The paper is organized as follows. Observations and data reductions are described in Section \ref{sect:obs}. Section \ref{sect:LV} presents the UV and optical light and color curves, while Section \ref{sect:Spe_analy} presents the spectral evolution. In Section \ref{sect:Discu} we constructed the bolometric luminosity of SN 2012fr and discussed its classification. The conclusions are given in Section \ref{sect:con}.

\section{Observations and Data Reduction}
\label{sect:obs}
\begin{figure}[!th]
\centering
\includegraphics[height=7cm,angle=0]{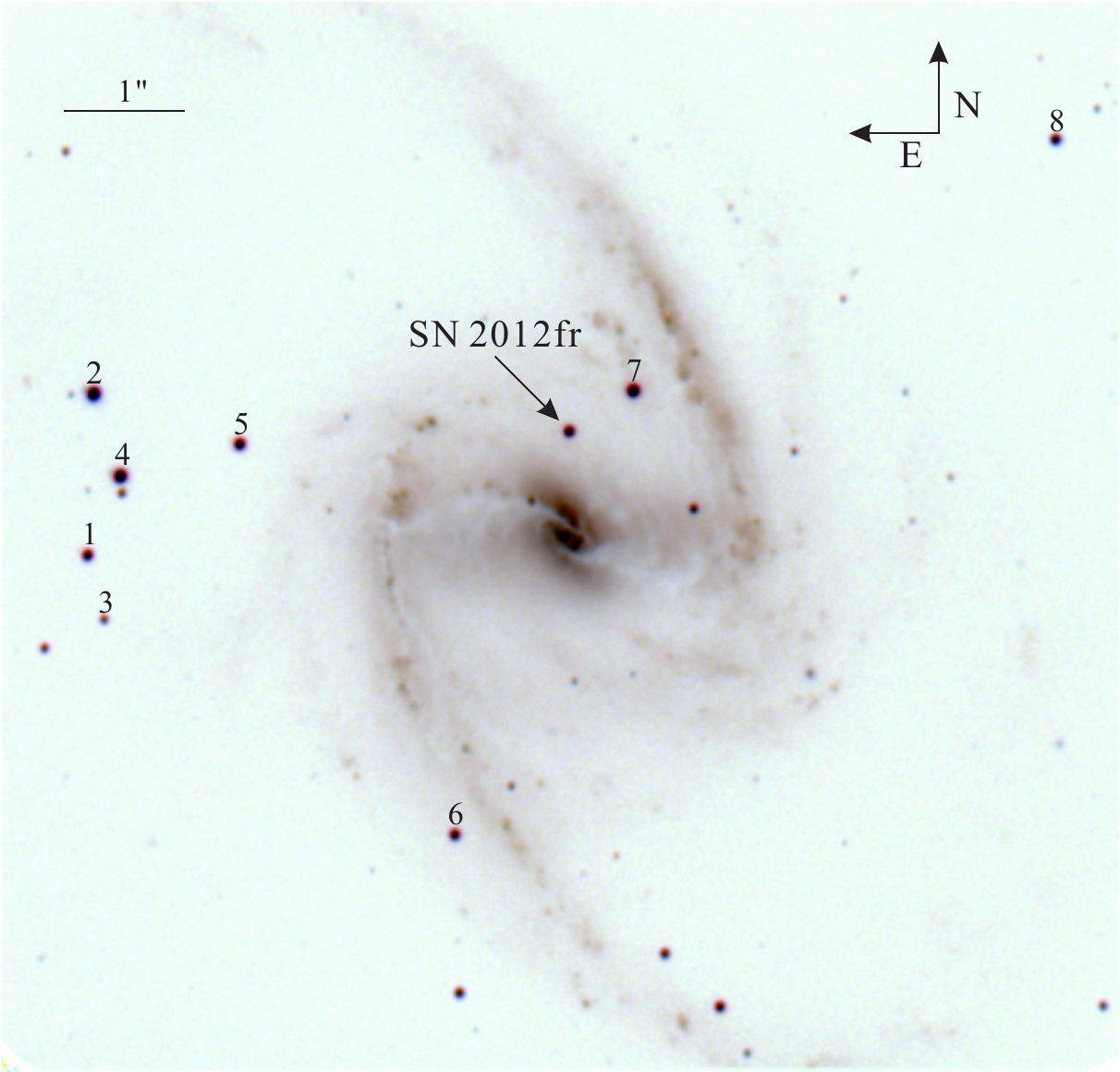}
\caption{SN 2012fr in NGC 1365. Composite color image obtained with the Li-Jiang 2.4-m telescope.
The supernova and eight local reference stars are marked.}
\label{<SN_image>}
\end{figure}

SN 2012fr was discovered by \citet{Klotz} on 2012 Oct. 27.05 (UT is used throughout this paper) using the 0.25~m robotic telescope TAROT La Silla observatory, Chile. Its coordinates are R.A. = 03h33m35s.99, Decl. = $-36^\circ$07\arcmin37.7\arcsec (J2000), located at 3\arcsec~west and 52\arcsec~north of the center of the nearby galaxy NGC 1365 (see Figure \ref{<SN_image>}). The host is a faced-on, Sb-type galaxy with a giant bar in the Fornax cluster, and has been extensively studied because of harboring a Seyfert 1.8 nucleus (see a review by Lindblad 1999). The supernova is located in a faint region that is not far from the bar of the host galaxy (see Figure 1).

An optical spectrum taken one day after the discovery shows that SN 2012fr was a young SN Ia (Childress et al. 2012), with extremely HVFs of \ion{Si}{2} and \ion{Ca}{2}. Our observations of this supernova started from 2012 Nov. 04 with the Yunnan Faint Object Spectrograph and Camera (YFOSC) mounted on the Li-Jiang 2.4-m telescope (hereafter LJT) of Yunnan Astronomical Observatories (YNAO), China. The YFOSC observation system is equipped with a 2.1K$\times$4.5K back-illuminated, blue sensitive CCD, which works in both the imaging and long-slit spectroscopic modes (see Zhang et al. 2012 for detailed descriptions of the YFOSC). In the imaging mode, the CCD has a field of view of $9\arcmin.6\times 9\arcmin.6$ (corresponding to an angular resolution of $0\arcsec.28$ per pixel). With this system, we collected a total of 42 photometric datapoints and 20 spectra in the optical. Extensive UV and optical photometry were also obtained with the Ultraviolet/Optical Telescope (UVOT) on the space-based $Swift$ telescope.

\subsection{Photometry}
\label{subsect:Phobs}
\subsubsection{Optical Observations from Li-Jiang 2.4-m Telescope}
\label{subsubsect:Ph_2m4}
Our optical photometry for SN 2012fr was obtained in the $BVRI$ bands with the LJT and YFOSC system, covering the period from t = $\sim$+11 days to t =$\sim$+71 days since the $B$-band maximum light. All of the CCD images were reduced using the IRAF \footnote{IRAF, the Image Reduction and Analysis Facility, is distributed by the National Optical Astronomy Observatory, which is operated by the Association of Universities for Research in Astronomy(AURA), Inc. under cooperative agreement with the National Science Foundation(NSF).} standard procedure, including the corrections for bias, flat field, and removal of cosmic rays. As SN 2012fr was very bright and in a faint region of the host galaxy (see Figure 1), we skipped the step of subtracting the galaxy template in the photometry. The standard point-spread function (PSF) fitting method from the IRAF DAOPHOT package \citep{Stetson} was used to measure the instrumental magnitudes for both the SN and the local standard stars. These magnitudes were then converted to the standard Johnson $UBV$ \citep{Johnson} and Kron-Cousins $RI$ \citep{Cousins} systems through transformations established by observing \citet{Landolt} standards during photometric nights. The average value of the photometric zeropoints, determined on four photometric nights, was used to calibrate the local standard stars in the field of SN 2012fr. Table \ref{Tab:Photo_stand} lists the standard $BVRI$ magnitudes and the corresponding uncertainties of 8 local standard stars labeled in Figure \ref{<SN_image>}. The magnitudes of these stars are then used to transform the instrumental magnitudes of SN 2012fr to those of the standard $BVRI$ system, and the final results of the photometry from the LJT are listed in Table \ref{Tab:2m4pho}.

\begin{deluxetable}{ccccc}[!th]
\tablecaption{Magnitudes of the Photometric Standards in the Field of SN 2012fr}
\tablewidth{0pt}

\tablehead{\colhead{Star} & \colhead{$B$(mag)} & \colhead{$V$(mag)} & \colhead{$R$(mag)} & \colhead{$I$(mag)} }
\startdata
1 & 15.59(04) & 15.04(03) & 14.61(02) & 14.24(03) \\
2 & 14.71(02) & 14.39(04) & 12.54(02) & 11.86(04) \\
3 & 17.31(05) & 16.66(05) & 16.21(03) & 15.76(02) \\
4 & 14.22(04) & 13.56(04) & 13.07(01) & 13.55(01) \\
5 & 14.89(03) & 14.57(03) & 14.26(02) & 13.97(05) \\
6 & 15.52(04) & 15.12(03) & 14.74(03) & 14.39(06) \\
7 & 14.06(04) & 13.59(02) & 13.16(03) & 12.76(04) \\
8 & 16.42(02) & 15.32(03) & 14.46(03) & 13.77(04) \\
\enddata
\tablecomments{See Figure 1 for the finder chart of SN 2012fr and the comparison stars.}
\tablecomments{Uncertainties, in units of 0.01 mag, are 1 $\sigma$.}
\label{Tab:Photo_stand}

\end{deluxetable}

\begin{deluxetable}{cccccccccc}
\newpage
\tablewidth{0pt}
\tablecaption{The $BVRI$ Photometry of SN 2012fr from Li-Jiang 2.4-m telescope and YFOSC. }
\tablehead{\colhead{MJD} & \colhead{Day\tablenotemark{a}} & \colhead{$B$(mag)} & \colhead{$V$(mag)} & \colhead{$R$(mag)} & \colhead{$I$(mag)} }
\startdata
56254.18	&	10.68	&	12.52(03)	&	12.21(01)	&	12.21(03)	&	12.65(02)	\\
56256.20	&	12.70	&	12.69(03)	&	12.34(02)	&	12.35(02)	&	12.81(03)	\\
56260.17	&	16.67	&	13.06(02)	&	12.57(02)	&	12.52(03)	&	12.87(03)	\\
56266.17	&	22.67	&	13.53(02)	&	12.83(01)	&	12.70(01)	&	\nodata	\\
56273.14	&	29.64	&	14.12(03)	&	13.08(01)	&	12.81(02)	&	12.57(04)	\\
56278.11	&	34.61	&	14.47(04)	&	13.37(01)	&	13.05(02)	&	12.77(03)	\\
56280.13	&	36.63	&	14.64(03)	&	13.54(01)	&	13.16(01)	&	12.91(04)	\\
56290.10	&	46.60	&	\nodata	   &	14.04(02)	&	13.76(02)	&	13.50(03)	\\
56298.10	&	54.60	&	14.99(04)	&	14.26(01)	&	14.00(02)	&	13.91(04)	\\
56306.05	&	62.55	&	15.09(03)	&	14.41(01)	&	14.28(02)	&	14.27(03)	\\
56314.04	&	70.54	&	15.19(04)	&	14.71(02)	&	14.52(02)	&	14.58(03)	\\
\enddata
\tablecomments{Uncertainties, in units of 0.01 mag, are 1$\sigma$; MJD = JD-2400000.}
\tablenotetext{a}{Relative to the epoch of $B$-band maximum (MJD =56243.50) }
\label{Tab:2m4pho}
\end{deluxetable}

\subsubsection{Optical/UV Observations from $Swfit$ UVOT}
\label{subsubsect:Ph_swift}
SN 2012fr was also intensively observed with the UVOT \citep{UVOT05} on board the $Swift$ satellite \citep{Swift04}, spanning from $t \approx -14$ days to $t \approx +126$ days relative to the $B$-band maximum light. The photometric observations were performed in three UV filters ($uvw2$, $uvm2$, and $uvw1$) and three broadband optical filters ($U$,$B$ and $V$). All of the $Swift$ images were reduced using HEASoft (the High Energy Astrophysics Software)\footnote{http://www.swift.ac.uk/analysis/software.php}-- a ``uvotsource" program with the latest $Swift$ Calibration Database\footnote{http://heasarc.gsfc.nasa.gov/docs/heasarc/caldb/swift/}. As the PSF profile is slightly dependent on the count rate of the source,  we adopted the aperture photometry in the reduction. The photometric aperture and the background value were set according to \citet{UVOTcali}. Since the UVOT is a photon-counting detector and suffers from coincidence loss (C-loss) for bright sources, the observed counts need corrections for such losses. This has now been automatically done by the ``uvotsource" program based on an empirical relation given by Poole et~al. (2008). As the instrumental response curves of the UVOT optical filters do not follow exactly those of the Johnson $UBV$ system, color-term corrections have been further applied to the magnitudes. Table \ref{Tab:Swiftpho} lists the final UVOT UV/optical magnitudes of SN 2012fr.

\subsection{Spectroscopy}
\label{sect:Sp_obs}
A total of 20 low-resolution spectra of SN 2012fr were obtained with the YFOSC on the LJT, with the wavelength covering from $\sim$3400\AA~ to $\sim$9100\AA. A few very early-time spectra obtained through Astronomical Ring for Access to Spectroscopy (ARAS\footnote{http://www.astrosurf.com/aras}) were also included in our analysis. A journal of our spectroscopic observations is given in Table \ref{Tab:Spec_log}.

All spectra were reduced using standard IRAF routines. The spectra were flux-calibrated with the spectrophotometric flux standard stars (e.g., LTT-1020) observed at a similar air mass on the same night. The spectra were further corrected for the continuum atmospheric extinction at the Li-Jiang Observatory; moreover, telluric lines were also removed from the data.

\begin{figure}[!th]
\centering
\includegraphics[width=8cm,angle=0]{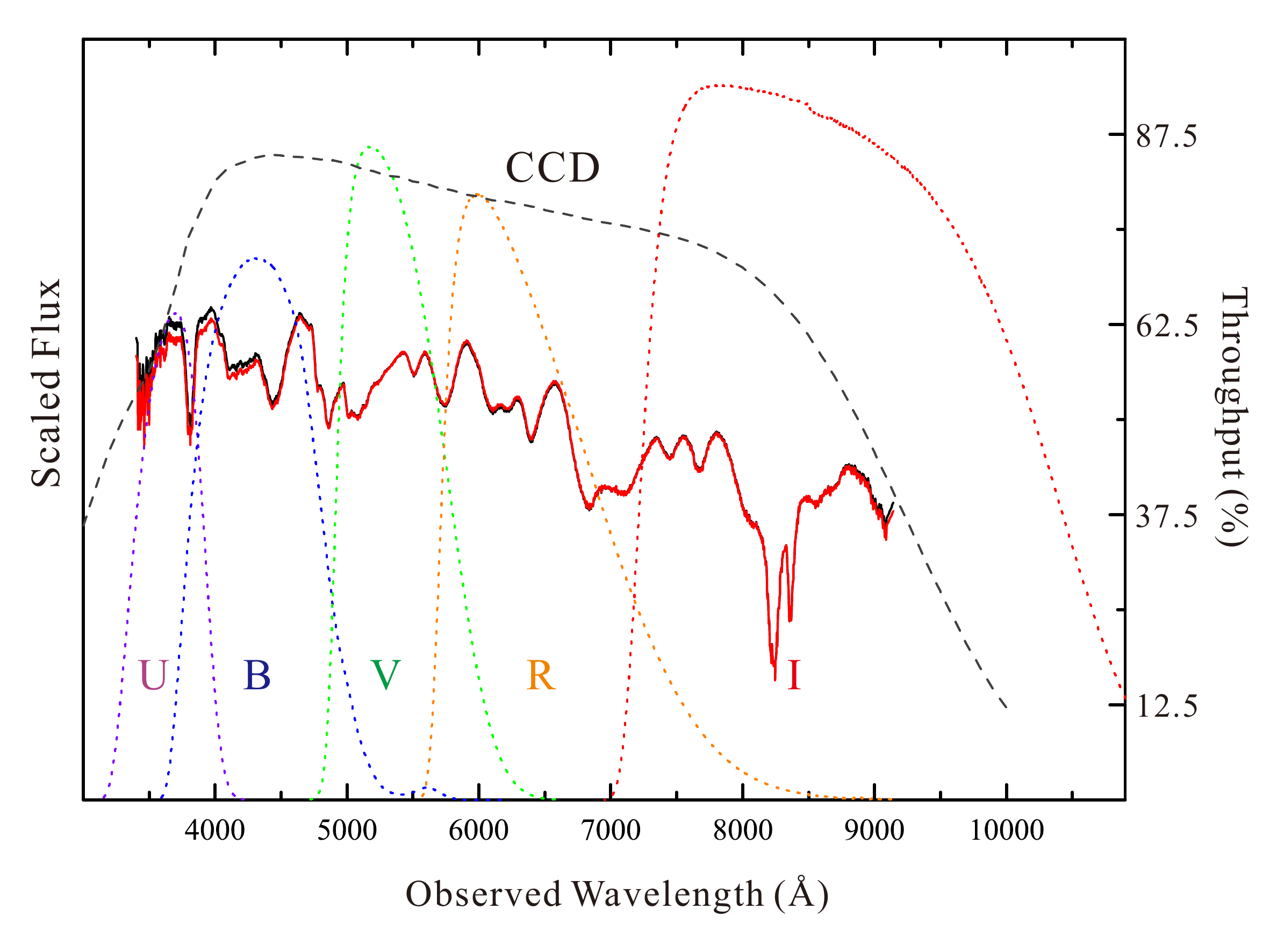}
\caption{A spectrum of SN 2012fr taken on Jan. 13 2013, with the flux calibrated by the spectrophotometric star LTT-1020 (black) and
another two standard stars of GD71 and G193-74 (red).}
\label{<Mul_stand>}
\end{figure}

\begin{deluxetable*}{cccccccc}[!th]
\tablewidth{0pt}
\tablecaption{$Swift$ UVOT Photometry of SN 2012fr}

\tablehead{\colhead{MJD} & \colhead{Day\tablenotemark{a}} & \colhead{uvw2(mag)} & \colhead{uvm2(mag)} & \colhead{uvw1(mag)} & \colhead{$U$(mag)} & \colhead{$B$(mag)} & \colhead{$V$(mag)} }

\startdata
56229.28	&	-14.22	&	19.19(21)	&	20.49(34)	&	17.50(12)	&	15.93(05)	&	15.21(03)	&	14.88(04)	\\
56229.29	&	-14.21	&	19.04(20)	&	20.32(22)	&	17.83(14)	&	15.83(05)	&	15.15(03)	&	14.91(04)	\\
56231.56	&	-11.94	&	17.14(09)	&	19.36(22)	&	15.62(08)	&	13.76(03)	&	13.62(03)	&	13.62(03)	\\
56232.08	&	-11.42	&	16.83(08)	&	18.81(15)	&	15.21(08)	&	13.43(03)	&	13.38(03)	&	13.42(03)	\\
56232.35	&	-11.15	&	16.73(08)	&	18.44(12)	&	15.07(07)	&	13.29(03)	&	13.30(03)	&	13.34(03)	\\
56232.42	&	-11.08	&	16.73(08)	&	18.58(13)	&	14.97(07)	&	13.25(03)	&	13.26(03)	&	13.31(03)	\\
56233.90	&	-9.60	&	15.97(07)	&	17.63(14)	&	14.18(07)	&	12.59(03)	&	12.85(03)	&	12.88(03)	\\
56233.96	&	-9.54	&	15.96(07)	&	17.64(10)	&	14.18(07)	&	12.59(03)	&	12.80(03)	&	12.88(03)	\\
56235.58	&	-7.92	&	15.45(04)	&	17.14(05)	&	13.74(03)	&	12.15(03)	&	12.65(03)	&	12.57(01)	\\
56235.64	&	-7.86	&	15.54(08)	&	16.88(10)	&	13.87(08)	&	12.24(04)	&	\nodata	&	12.58(04)	\\
56235.71	&	-7.79	&	15.52(08)	&	17.01(11)	&	13.61(08)	&	12.13(04)	&	\nodata	&	12.55(04)	\\
56235.78	&	-7.72	&	15.48(10)	&	16.95(11)	&	13.67(08)	&	12.16(04)	&	\nodata	&	12.53(04)	\\
56235.91	&	-7.59	&	15.53(08)	&	17.80(16)	&	13.57(03)	&	12.18(05)	&	\nodata	&	12.60(04)	\\
56237.70	&	-5.80	&	15.07(04)	&	16.56(05)	&	13.27(03)	&	11.97(03)	&	12.40(03)	&	12.34(01)	\\
56237.90	&	-5.60	&	15.21(07)	&	16.65(09)	&	13.41(07)	&	\nodata	&	\nodata	&	12.32(04)	\\
56239.57	&	-3.93	&	14.92(04)	&	16.21(05)	&	13.11(03)	&	11.79(03)	&	12.14(03)	&	12.15(01)	\\
56241.64	&	-1.86	&	14.88(03)	&	15.96(05)	&	13.12(03)	&	11.74(03)	&	12.09(03)	&	12.06(01)	\\
56244.18	&	+0.68	&	14.94(03)	&	15.82(05)	&	13.29(03)	&	11.85(01)	&	12.03(03)	&	12.01(01)	\\
56246.32	&	+2.82	&	\nodata	&	\nodata	&	\nodata	&	12.02(02)	&	12.08(02)	&	\nodata	\\
56248.06	&	+4.56	&	15.17(03)	&	15.87(04)	&	13.70(03)	&	12.12(01)	&	12.15(03)	&	12.05(01)	\\
56248.38	&	+4.88	&	\nodata	&	\nodata	&	\nodata	&	12.15(06)	&	12.18(03)	&	\nodata	\\
56249.52	&	+6.02	&	\nodata	&	\nodata	&	\nodata	&	12.22(02)	&	12.24(02)	&	12.13(05)	\\
56250.13	&	+6.63	&	15.21(03)	&	15.89(04)	&	13.72(07)	&	12.25(03)	&	\nodata	&	12.10(03)	\\
56251.54	&	+8.04	&	\nodata	&	\nodata	&	\nodata	&	12.38(02)	&	12.36(02)	&	12.11(05)	\\
56251.66	&	+8.16	&	15.33(05)	&	15.98(06)	&	13.93(08)	&	12.36(04)	&	\nodata	&	12.12(04)	\\
56251.73	&	+8.23	&	15.38(04)	&	15.94(05)	&	13.91(07)	&	12.44(04)	&	\nodata	&	12.13(03)	\\
56254.00	&	+10.50	&	15.54(04)	&	16.17(05)	&	14.13(07)	&	12.68(03)	&	12.58(04)	&	12.26(03)	\\
56256.35	&	+12.85	&	\nodata	&	\nodata	&	\nodata	&	12.94(03)	&	12.84(06)	&	12.43(05)	\\
56256.41	&	+12.91	&	15.80(04)	&	16.48(05)	&	14.50(07)	&	12.90(03)	&	12.79(03)	&	12.41(03)	\\
56257.88	&	+14.38	&	16.03(04)	&	16.63(06)	&	14.66(07)	&	13.07(03)	&	12.89(03)	&	12.49(03)	\\
56257.95	&	+14.45	&	\nodata	&	\nodata	&	\nodata	&	13.07(05)	&	13.02(05)	&	12.46(05)	\\
56260.21	&	+16.71	&	16.27(05)	&	16.86(07)	&	14.90(08)	&	13.38(03)	&	13.12(03)	&	12.57(03)	\\
56260.27	&	+16.77	&	\nodata	&	\nodata	&	\nodata	&	13.36(03)	&	13.12(05)	&	12.61(05)	\\
56262.35	&	+18.85	&	16.55(05)	&	17.12(07)	&	15.14(08)	&	13.60(03)	&	13.28(03)	&	12.75(03)	\\
56262.42	&	+18.92	&	\nodata	&	\nodata	&	\nodata	&	13.61(03)	&	13.32(05)	&	12.70(02)	\\
56264.08	&	+20.58	&	\nodata	&	\nodata	&	\nodata	&	13.70(06)	&	13.43(02)	&	12.73(03)	\\
56264.15	&	+20.65	&	16.69(07)	&	17.20(09)	&	15.37(08)	&	13.73(03)	&	13.46(03)	&	12.78(03)	\\
56266.22	&	+22.72	&	\nodata	&	\nodata	&	\nodata	&	13.99(03)	&	13.61(02)	&	12.83(05)	\\
56266.29	&	+22.79	&	16.77(06)	&	17.30(08)	&	15.61(08)	&	14.01(03)	&	13.60(03)	&	12.84(03)	\\
56267.51	&	+24.01	&	\nodata	&	\nodata	&	\nodata	&	14.16(03)	&	13.86(05)	&	12.96(06)	\\
56268.23	&	+24.73	&	17.12(07)	&	17.54(09)	&	15.73(08)	&	14.17(04)	&	13.79(02)	&	12.91(03)	\\
56269.63	&	+26.13	&	\nodata	&	\nodata	&	\nodata	&	14.30(03)	&	13.91(05)	&	12.94(05)	\\
56270.09	&	+26.59	&	17.03(07)	&	17.60(10)	&	15.88(09)	&	14.28(04)	&	13.93(03)	&	12.94(03)	\\
56275.63	&	+32.13	&	17.36(08)	&	17.74(10)	&	16.23(09)	&	14.66(04)	&	14.33(03)	&	13.24(03)	\\
56280.18	&	+36.68	&	17.52(08)	&	18.03(10)	&	16.89(10)	&	14.96(04)	&	14.63(03)	&	13.53(03)	\\
56283.51	&	+40.01	&	17.60(11)	&	18.21(16)	&	16.85(15)	&	15.18(05)	&	14.75(04)	&	13.68(04)	\\
56287.92	&	+44.42	&	17.87(11)	&	18.22(14)	&	16.97(14)	&	15.24(05)	&	14.84(03)	&	13.88(04)	\\
56291.67	& 	+48.17	&	18.02(11)	&	18.31(12)	&	17.13(13)	&	15.27(04)	&	14.93(03)	&	14.03(03)	\\
56296.01	&	+52.51	&	18.11(11)	&	18.53(15)	&	17.18(14)	&	15.34(04)	&	14.97(03)	&	14.21(04)	\\
56300.22	&	+57.22	&	18.36(13)	&	18.48(13)	&	17.24(13)	&	15.48(04)	&	15.01(03)	&	14.33(04)	\\
56303.69	&	+60.69	&	18.12(17)	&	18.51(18)	&	17.32(15)	&	15.52(07)	&	15.05(04)	&	14.39(05)	\\
56314.65	&	+71.65	&	18.61(17)	&	18.53(15)	&	17.47(17)	&	15.69(06)	&	15.20(04)	&	14.65(05)	\\
56320.98	&	+77.98	&	18.62(20)	&	18.99(31)	&	17.65(20)	&	15.81(05)	&	15.37(03)	&	14.81(04)	\\
56326.66	&	+83.66	&	18.71(20)	&	18.96(26)	&	17.64(19)	&	15.85(07)	&	15.43(05)	&	14.98(06)	\\
56332.75	&	+89.75	&	18.68(16)	&	19.08(29)	&	17.62(23)	&	15.91(06)	&	15.48(06)	&	15.02(05)	\\
56339.02	&	+96.02	&	18.83(18)	&	19.21(22)	&	17.63(16)	&	16.08(06)	&	15.61(05)	&	15.34(06)	\\
56344.50	&	+101.50	&	18.98(20)	&	19.52(28)	&	17.91(19)	&	16.26(07)	&	15.73(04)	&	15.43(06)	\\
56350.64	&	+107.64	&	19.00(20)	&	19.50(26)	&	18.00(26)	&	16.30(07)	&	15.81(07)	&	15.62(06)	\\
56356.65	&	+113.65	&	19.03(18)	&	19.41(21)	&	18.19(24)	&	16.37(07)	&	15.86(04)	&	15.73(06)	\\
56362.53	&	+119.53	&	19.08(18)	&	19.72(32)	&	18.21(21)	&	16.54(09)	&	15.95(05)	&	15.83(07)	\\
56368.95	&	+125.95	&	19.12(21)	&	19.94(34)	&	18.41(21)	&	16.66(09)	&	16.09(05)	&	15.85(07)	\\
\enddata
\tablecomments{Uncertainties, in units of 0.01 mag, are 1$\sigma$; MJD = JD-2400000.}
\tablenotetext{a}{Relative to the epoch of $B$-band Maximum}
\label{Tab:Swiftpho}
\end{deluxetable*}

\subsection{Spectrophotometry}
\label{subsect:spc_ph}
As it can be seen that the spectra obtained with the LJT for SN 2012fr have a better phase coverage compared with the photometry. In particular, these spectra have better flux calibrations and cover well the wavelength range of the $BVR$ filters, as shown in Figure \ref{<Mul_stand>}. This enables us to fill in the gaps of the light curves through spectrophotometry. The synthetic magnitudes were derived by convolving the observed spectra with the transmission curves of the filters, and are listed in Table \ref{Tab:Spe2mag}. The spectrophotometry is overall consistent with the photometry to within 0.10 mag in the $BVR$ bands, but larger differences emerge in the $I$ band because of limited wavelength coverage of our spectra. The errors listed in Table \ref{Tab:Spe2mag} include uncertainties in slit loss, sky transparency loss, and flux calibration of the spectra. A slit width of $1\arcsec.8$ was used in the observations, which is larger than the typical seeing at Li-Jiang observatory (i.e., $\sim$ 1$\arcsec$). The slit-loss is therefore negligible for our spectra. The uncertainty due to the sky transparency loss should be small as the flux standard star LTT-1020 is very close to SN 2012fr. In the $I$ band, the synthetic magnitudes still suffer from the uncertainty in the adopted spectral shape of the wavelength region from 9100\AA~to 11000\AA. A spectral template with a similar $\Delta m_{15}(B)$ was taken from Hsiao et al. (2007), and an uncertainty of 10\% was assumed in the calculation. Note that the uncertainties on the derived magnitudes may be underestimated as compared to those listed in Table \ref{Tab:Spe2mag} because of large airmasses through which SN 2012fr and LTT-1020 were observed at Li-Jiang Observatory.

\begin{deluxetable}{cccccc}

\tablewidth{0pt}
\tablecaption{Magnitudes of the Spectrophotometry of SN 2012fr}

\tablehead{\colhead{MJD} & \colhead{Day\tablenotemark{a}} & \colhead{$B$(mag)} & \colhead{$V$(mag)} & \colhead{$R$(mag)} & 
\colhead{$I$(mag)}}
\startdata
236.25	&	-7.25	&	12.44(10)	&	12.66(08)	&	12.30(08)	&	12.62(10)	\\
237.21	&	-6.29	&	12.30(10)	&	12.52(08)	&	12.23(08)	&	12.48(10)	\\
239.22	&	-4.28	&	12.11(10)	&	12.24(08)	&	12.05(08)	&	12.37(10)	\\
241.22	&	-2.28	&	12.02(20)	&	12.14(08)	&	11.93(08)	&	12.28(10)	\\
243.23	&	-0.27	&	11.95(10)	&	12.04(08)	&	11.89(08)	&	12.29(10)	\\
247.22	&	+3.72	&	12.08(10)	&	11.95(08)	&	11.97(08)	&	12.30(10)	\\
250.24	&	+6.74	&	12.18(10)	&	12.05(08)	&	12.12(08)	&	12.37(10)	\\
252.22	&	+8.72	&	12.42(10)	&	12.21(08)	&	12.21(08)	&	12.54(10)	\\
254.18	&	+10.68	&	12.51(10)	&	12.29(08)	&	12.22(08)	&	12.64(10)	\\
256.20	&	+12.70	&	12.65(10)	&	12.40(08)	&	12.38(08)	&	12.82(10)	\\
260.17	&	+16.67	&	12.99(10)	&	12.60(08)	&	12.59(08)	&	12.99(10)	\\
266.17	&	+22.67	&	13.59(10)	&	12.71(08)	&	12.63(08)	&	12.85(10)	\\
273.14	&	+29.64	&	14.17(10)	&	13.04(08)	&	12.79(08)	&	12.62(10)	\\
277.11	&	+33.61	&	14.32(20)	&	13.16(08)	&	12.97(08)	&	12.78(10)	\\
280.13	&	+36.63	&	14.44(10)	&	13.48(08)	&	13.12(08)	&	12.83(10)	\\
290.10	&	+46.60	&	14.81(10)	&	14.00(08)	&	13.67(08)	&	13.49(10)	\\
298.10	&	+54.60	&	14.96(10)	&	14.19(08)	&	13.89(08)	&	13.86(10)	\\
306.05	&	+62.55	&	15.11(10)	&	14.49(08)	&	14.21(08)	&	14.20(10)	\\
314.04	&	+70.54	&	15.25(10)	&	14.63(08)	&	14.44(08)	&	14.56(10)	\\
\enddata
\tablecomments{Uncertainties, in units of 0.01 mag, are 1$\sigma$; MJD = JD-2456000.
The spectra have been corrected for reddening and redshift before
being used for spectrophotometry. }
\tablenotetext{a}{Relative to the epoch of $B$-band Maximum}
\label{Tab:Spe2mag}
\end{deluxetable}

\begin{deluxetable*}{cccccccccc}[!th]

\tablewidth{0pt}
\tablecaption{Journal of Spectroscopic Observations of SN 2012fr}

\tabletypesize{\scriptsize}
\tablehead{\colhead{UT Date}&\colhead{MJD} & \colhead{Epoch} & \colhead{Res} & \colhead{Range} & \colhead{Telescope} & \colhead{Flux} & \colhead{Airmass} & \colhead{Exp.time} & \colhead{Observer} \\
\colhead{}&\colhead{(-240000)} & \colhead{(days)} & \colhead{(\AA/pix)} & \colhead{(\AA)} & \colhead{(+Instrument)} & \colhead{(Standard)} & \colhead{} & \colhead{(sec)} & \colhead{} }
\startdata
2012 Oct. 29	&	56230.07	&	-13.43	&	4.65	&	4000-6500	&	C11+Lhires	&	\nodata	&	1.05	&	7$\times$600	&	 BH\tablenotemark{a}	 \\
2012 Oct. 30	&	56230.99	&	-12.51	&	6.67	&	4000-7000	&	C11+LISA	&	HD 27376	&	1.22	&	3$\times$600	&	 TB\tablenotemark{b}	\\
2012 Oct. 31	&	56231.97	&	-11.53	&	6.67	&	4000-7000	&	C11+LISA	&	HD 27376	&	1.33	&	8$\times$600	&	TB	\\
2012 Nov. 4	&	56236.25	&	-7.25	&	2.85	&	3400-9100	&	LJT+YFOSC	&	LTT1020	&	2.25	&	600.00	&	ZJ\tablenotemark{c}	\\
2012 Nov. 5	&	56237.21	&	-6.29	&	2.85	&	3400-9100	&	LJT+YFOSC	&	LTT1020	&	2.17	&	600.00	&	ZJ	\\
2012 Nov. 7	&	56239.22	&	-4.28	&	2.85	&	3400-9100	&	LJT+YFOSC	&	LTT1020	&	2.18	&	600.00	&	ZJ	\\
2012 Nov. 9	&	56241.22	&	-2.28	&	2.85	&	3400-9100	&	LJT+YFOSC	&	LTT1020	&	2.19	&	600.00	&	ZJ	\\
2012 Nov. 11	&	56243.23	&	-0.27	&	2.85	&	3400-9100	&	LJT+YFOSC	&	LTT1020	&	2.06	&	600.00	&	ZJ	\\
2012 Nov. 14	&	56246.24	&	+2.74	&	2.85	&	3400-9100	&	LJT+YFOSC	&	\nodata\tablenotemark{d}	&	2.31	&	600.00	&	ZJ	 \\
2012 Nov. 15	&	56247.22	&	+3.72	&	2.85	&	3400-9100	&	LJT+YFOSC	&	LTT1020	&	2.23	&	600.00	&	ZJ	\\
2012 Nov. 18	&	56250.24	&	+6.74	&	2.85	&	3400-9100	&	LJT+YFOSC	&	LTT1020	&	2.42	&	600.00	&	ZJ	\\
2012 Nov.20	&	56252.22	&	+8.72	&	2.85	&	3400-9100	&	LJT+YFOSC	&	LTT1020	&	2.30	&	600.00	&	ZJ	\\
2012 Nov.22	&	56254.18	&	+10.68	&	2.85	&	3400-9100	&	LJT+YFOSC	&	LTT1020	&	2.17	&	600.00	&	ZJ	\\
2012 Nov.24	&	56256.20	&	+12.70	&	2.85	&	3400-9100	&	LJT+YFOSC	&	LTT1020	&	2.27	&	600.00	&	ZJ	\\
2012 Nov. 28	&	56260.17	&	+16.67	&	2.85	&	3400-9100	&	LJT+YFOSC	&	LTT1020	&	2.19	&	600.00	&	ZJ	\\
2012 Dec. 4	&	56266.17	&	+22.67	&	2.85	&	3400-9100	&	LJT+YFOSC	&	LTT1020	&	2.26	&	600.00	&	ZJ	\\
2012 Dec. 11	&	56273.14	&	+29.64	&	2.85	&	3400-9100	&	LJT+YFOSC	&	LTT1020	&	2.21	&	600.00	&	ZJ	\\
2012 Dec. 15	&	56277.11	&	+33.61	&	2.85	&	3400-9100	&	LJT+YFOSC	&	LTT1020	&	2.17	&	600.00	&	ZJ	\\
2012 Dec. 18	&	56280.13	&	+36.63	&	2.85	&	3400-9100	&	LJT+YFOSC	&	LTT1020	&	2.24	&	600.00	&	ZJ	\\
2012 Dec. 28	&	56290.10	&	+46.60	&	2.85	&	3400-9100	&	LJT+YFOSC	&	LTT1020	&	2.24	&	600.00	&	ZJ	\\
2013 Jan. 5	&	56298.10	&	+54.60	&	2.85	&	3400-9100	&	LJT+YFOSC	&	LTT1020	&	2.33	&	900.00	&	ZJ	\\
2013 Jan. 13	&	56306.05	&	+62.55	&	2.85	&	3400-9100	&	LJT+YFOSC	&	LTT1020	&	2.19	&	900.00	&	ZJ	\\
2013 Jan. 21	&	56314.04	&	+70.54	&	2.85	&	3400-9100	&	LJT+YFOSC	&	LTT1020	&	2.25	&	900.00	&	ZJ	\\
\enddata

\tablenotetext{a}{Bernard Heathcote. Facility: Celestron11 + LhiresIII150 + ATIK314L. Observatory: Melbourne (Australia)}
\tablenotetext{b}{Terry Bohlsen. Facility: Celestron11 + LISA ST8XME. Observatory: Armidale (Australia)}
\tablenotetext{c}{Zhang Jujia and staff of the Li-Jiang Observatory. Facility: LJT+YFOSC. Observatory: YNAO (China)}
\tablenotetext{d}{flux calibration was obtained through the observations
of LTT1020 on Nov. 13 and Nov. 15 2012.}
\label{Tab:Spec_log}
\end{deluxetable*}

\section{Light Curves of SN 2012fr}
\label{sect:LV}
\begin{figure}[!th]
\centering
\includegraphics[width=7cm,angle=0]{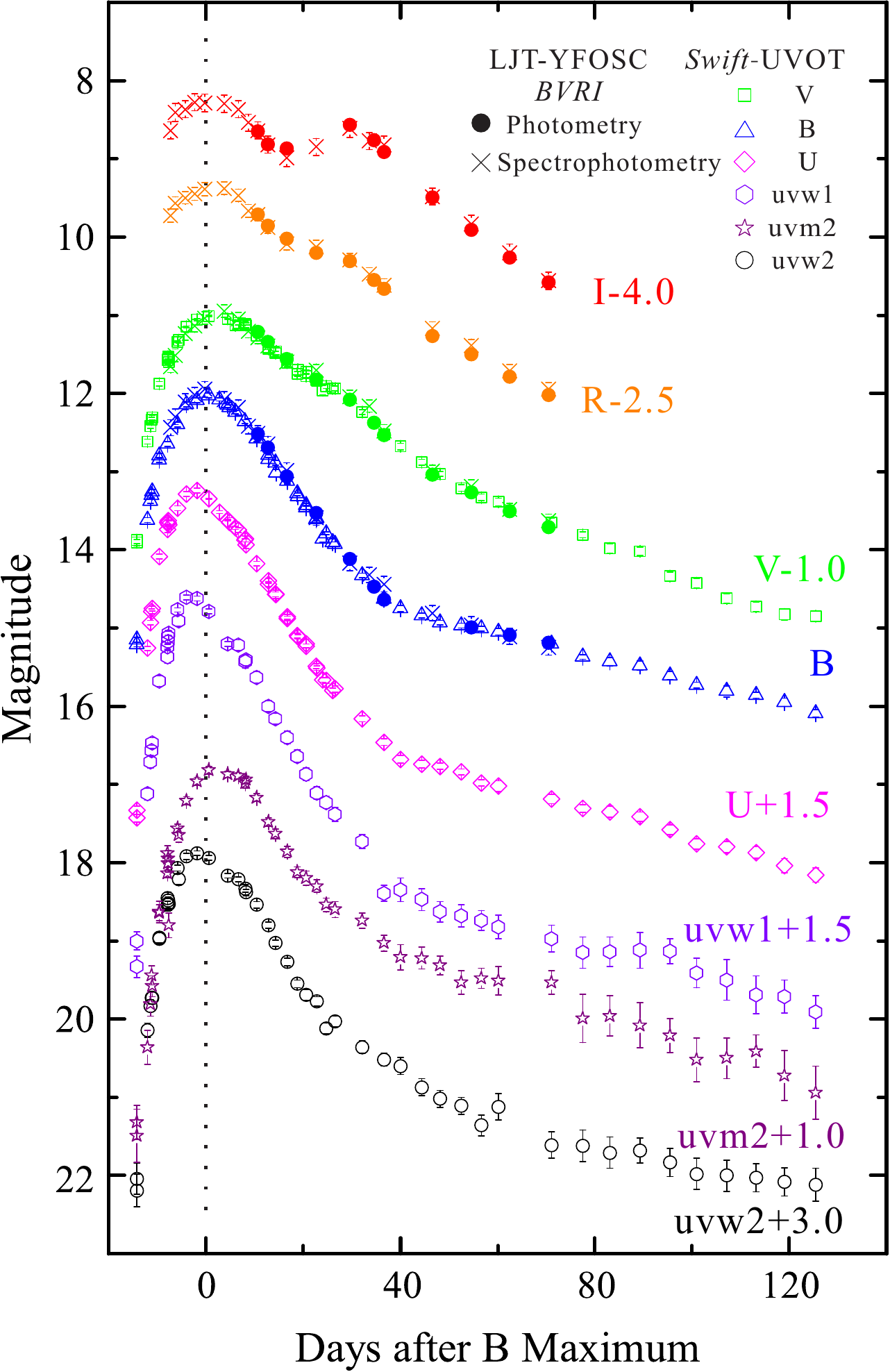}
\caption{Ultraviolet and optical light curves of SN 2012fr. The light curves are shifted vertically for better display.}
\label{<LightCurve>}
\end{figure}

Figure \ref{<LightCurve>} shows the optical and UV light curves of SN 2012fr. Besides the color-term corrections, additional magnitude corrections such as S-corrections (Stritzinger et al. 2002) are also applied to the light curves to account for the differences between the instrumental responses and those defined by Bessell (1992). No S-corrections were applied to the UV data. In the optical bands, the observations of the LJT and YFOSC system agree well with the $Swift$ observations. Detailed analysis of the light and color curves are presented in the following subsections.

\subsection{Optical and Ultraviolet Light Curves}
\label{subsect:OPLC}
With the LJT and $Swift$ light curves, we derived the parameters of peak magnitudes, maximum dates, and light-curve decline rates $\Delta m_{15}$ (i.e., Phillips 1993), as listed in Table \ref{Tab:LV_par}. It is found that SN 2012fr reached a $B$-band maximum brightness of $12.01\pm0.01$ mag on JD 2456243.50$\pm$0.30 (2012 Nov. 12.00 UT), and it reached at the $V$-band maximum of 11.99$\pm$0.01 mag at about 1.5 days later. The observed light-curve decline rate $\Delta m_{15}$(B) and the maximum-light color $B_{max} - V_{max}$ are estimated as 0.85$\pm$0.05 mag and $0.02\pm$0.03 mag, respectively.

 \begin{figure*}[!th]
\centering
\includegraphics[height=7.1cm,angle=0]{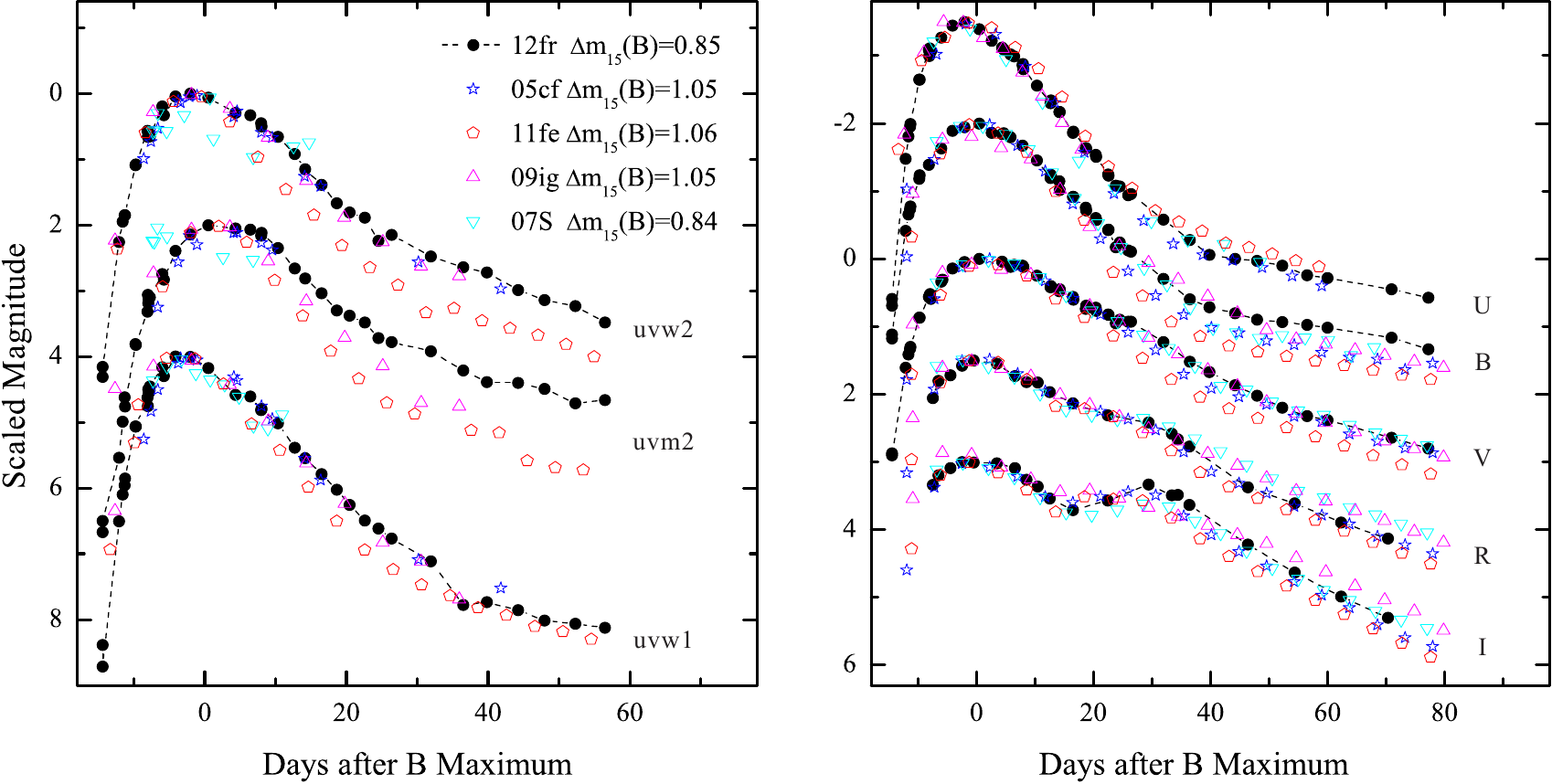}
\caption{Light curves of SN 2012fr compared with some representative sample of SNe Ia. Left: Comparison of SN 2012fr with SN 2005cf, SN 2007S, SN 2009ig, and SN 2011fe in $Swift$ UV bands (see text for references); Right: Comparison in the $UBVRI$ bands (see text for references).}
\label{<LCcomp>}
\end{figure*}

Our estimates of these parameters are consistent with those reported by C13 who derived that $B$-band light curve has a peak brightness of $\approx$12.0 mag on 2012 Nov. 12.04 and a decline rate $\Delta m _{15}$(B) = 0.80 mag. A smaller decline rate suggests that SN 2012fr should be intrinsically luminous if it follows the light curve width-luminosity relation (i.e., Phillips 1993; Riess et al. 1996; Goldhabor et al. 2001; Guy et al. 2005). After corrections for the Galactic reddening of E$(\bv)$ = 0.02 mag (Schlegel et al. 1998), the corresponding color index of $B_{max}-V_{max}$ becomes 0$\pm$0.03 mag. This suggests insignificant host-galaxy reddening for SN 2012fr according to the intrinsic colors of SNe Ia (e.g., Phillips et al. 1999; Wang et al. 2009a).

Figure \ref{<LCcomp>} shows the comparison of SN 2012fr with some well-observed SNe Ia with both optical and UV observations, including SN 2005cf ($\Delta m_{15}$ = 1.05 mag; Wang et al. 2009b), SN 2007S ($\Delta m_{15}$ = 0.84 mag; Brown et al. 2009), SN 2009ig ($\Delta m_{15}$ = 1.05 mag; Foley et al. 2012, and Marion et al. 2013), and SN 2011fe ($\Delta m_{15}$ = 1.06 mag; Brown et al. 2012, and Wang et al. 2014 in prep.). The sample selected for comparison have similar values of $\Delta m_{15}$ (B). Of this sample, SN 2005cf and SN 2011fe can be put into the NV (or LVG) subclass, while SN 2007S and SN 2009ig can be put into the 91T-like (or SS) subclass and HV subclass, respectively. In the UV bands, the light curve peaks of SN 2012fr are similar to that of the comparison sample except for SN 2011fe which has an apparently faster post-maximum decline rate. Difference also exists in the optical bands where the tail of SN 2011fe appears to be fainter than SN 2012fr and other comparison SNe by 0.3--0.5 mag, depending on the wavebands.

We further noticed that SN 2012fr perhaps has a more prominent shoulder feature at about 30 days after the maximum brightness in comparison with other SNe~Ia of our sample. This feature is stronger in SN 2012fr than the others in the $I$ band, bears about the same strength as the others in the $R$ band, and may be a little stronger than the others in the $V$ band. In observations, the relative strength of the $I$-band secondary shoulder is found to correlate with decline rate (and hence peak luminosity), being more prominent and occurring later in more luminous SNe Ia. This observed behavior may be taken as further evidence that SN 2012fr has a larger production of nickel (and hence iron group elements) in the explosion (Kasen 2006), consistent with the $^{56}$Ni estimated in Section 5.1.

In general, the overall light curve evolution of SN 2012fr resembles closely to those of SN 2007S and SN 2009ig, although SN 2007S is not well sampled in the UV bands. SN 2011fe matches well with SN 2012fr near the maximum phase, but they show noticeable differences in the later phases (especially at shorter wavelengths).

\begin{deluxetable}{ccccc}[!th]
\tablewidth{0pt}
\tablecaption{Light-Curve Parameters of SN 2012fr}

\tablehead{\colhead{Band} & \colhead{$\lambda_{central}$} & \colhead{t$_{max}$\tablenotemark{a}} & \colhead{m$_{peak}$\tablenotemark{b}} & \colhead{$\Delta$ m$_{15}$\tablenotemark{b}} \\
\colhead{} & \colhead{(\AA)} & \colhead{(-240000)} & \colhead{(mag)} & \colhead{(mag)} }

\startdata
uvw2 & 1928 & 56243.40(40) & 15.00(04) & 1.03(06) \\
uvm2 & 2246 & 56246.10(60) & 15.81(05) & 0.98(07) \\
uvw1 & 2600 & 56240.60(50) & 13.27(03) & 1.09(05) \\
U & 3650 & 56242.10(40) & 11.89(03) & 1.09(05) \\
B & 4450 & 56243.50(30) & 12.01(03) & 0.85(05) \\
V & 5500 & 56245.00(30) & 11.99(03) & 0.59(05) \\
R & 6450 & 56244.90(80) & 11.89(10) & 0.62(10) \\
I & 7870 & 56242.50(80) & 12.28(10) & 0.55(10) \\
\enddata

\tablenotetext{a}{Uncertainties of peak-light dates, in units of 0.01 day, are 1 $\sigma$}
\tablenotetext{b}{Uncertainties of magnitudes, in units of 0.01 mag, are 1 $\sigma$}

\label{Tab:LV_par}
\end{deluxetable}

\subsection{Color Curves and Interstellar Extinction}
\label{subsect:CV}
\begin{figure}
\centering
\includegraphics[width=8cm,angle=0]{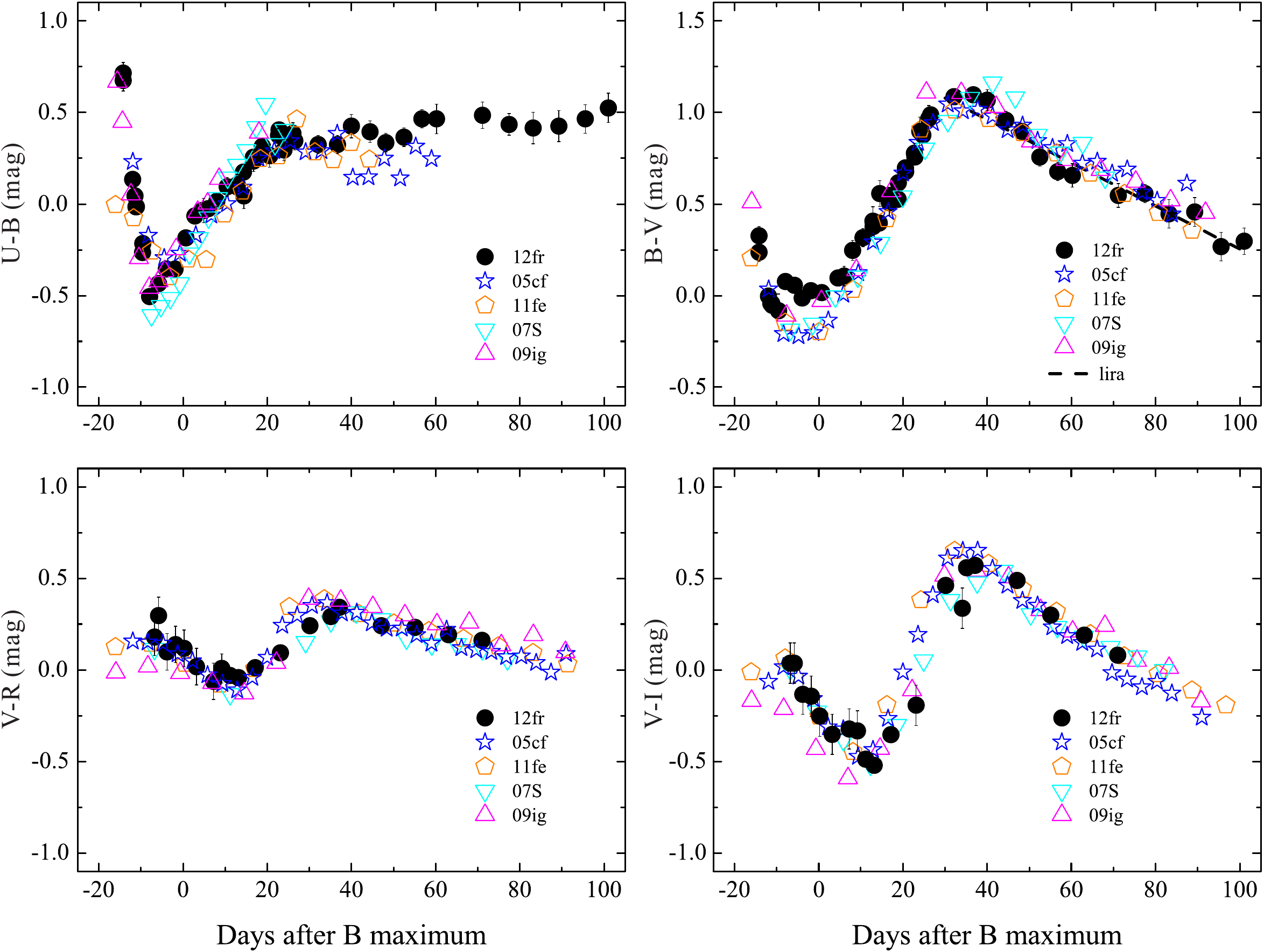}
\caption{Comparison of the color curves between SN 2012fr and some representative SNe Ia (see text for references).
The dashed line in the $\bv$ panel shows the $Lira-Phillips$ loci (Phillips et al. 1999).}
\label{<colCurve>}
\end{figure}

Figure \ref{<colCurve>} shows the color curves of SN 2012fr, corrected for the Galactic reddening from Schlegel et al. (1998) and the host-galaxy reddening derived from the following analysis. Overplotted are the color curves of SN 2005cf, SN 2007S, SN 2009ig, and SN 2011fe. The overall color evolution of SN 2012fr is similar to those of the selected SNe Ia, but scatter presents in the colors at shorter wavelengths (i.e., $U - B$ and $B - V$). In $U - B$, SN 2012fr reached at the bluest color at about one week before $B$-band maximum light, as similarly seen in SN 2009ig and SN 2007S. Of these sample, SN 2007S seems to have a relatively bluer color at this stage. SN 2005cf and SN 2011fe reached at their bluest colors a few days later and they are also redder relative to the other three sample. In $B - V$, SN 2012fr is redder than the comparison SNe Ia by $\sim$0.2 mag near the maximum light. During the period from t $\sim$ 40 days to t $\sim$ 80 days, the $B - V$ color of SN 2012fr appears somewhat bluer than that of SN 2005cf and SN 2011fe, with a slope steeper than the $Lira-Phillips$ relation (Phillips et al. 1999). A similar feature of faster $B - V$ evolution is also seen in the HV SNe Ia at comparable phases (e.g., Wang et al. 2008). The $V - R$ and $V - I$ color curves of SN 2012fr exhibits a behavior that is very similar to those of the comparison SNe.

The Galactic reddening of SN 2012fr is $E(\bv)_{Gal}$ = 0.02 mag \citep{Schle98}, corresponding to an extinction of 0.06 mag adopting the standard reddening law of \citet{Cardelli}. The reddening due to the host galaxy can be estimated by several empirical methods. A value of $\sim$0.1 mag can be derived for E$(B - V)_{host}$ by using the empirical relation established between the intrinsic $B_{max} - V_{max}$ color and $\Delta m_{15}$ (i.e., Phillips et al. 1999; Wang et al. 2009b). The comparison of the late-time $B - V$ color with that predicted by the $Lira-Phillips$ relation (see the dashed line in Figure 5) gives a host-galaxy reddening of E$(B - V)_{host}$ = $-$0.07 mag for SN 2012fr. The negative reddening is apparently unphysical. This inconsistency indicates that the intrinsic color is not well understood yet for SNe Ia of different subtypes; and any empirical method should be used with caution for any individual SN Ia. On the other hand, the high-resolution spectrum of SN 2012fr does not show any significant signature of \ion{Na}{1} D absorption feature (C13), suggesting a low reddening. C13 therefore placed an upper limit, with $E(\bv) < 0.015$, for the host-galaxy reddening of SN 2012fr. In the following analysis, we assume the total line-of-sight reddening of SN 2012fr as $E(\bv)$ = 0.03 mag.

\section{Optical Spectra}
\label{sect:Spe_analy}
\begin{figure}[!th]
\centering
\includegraphics[width=8cm,angle=0]{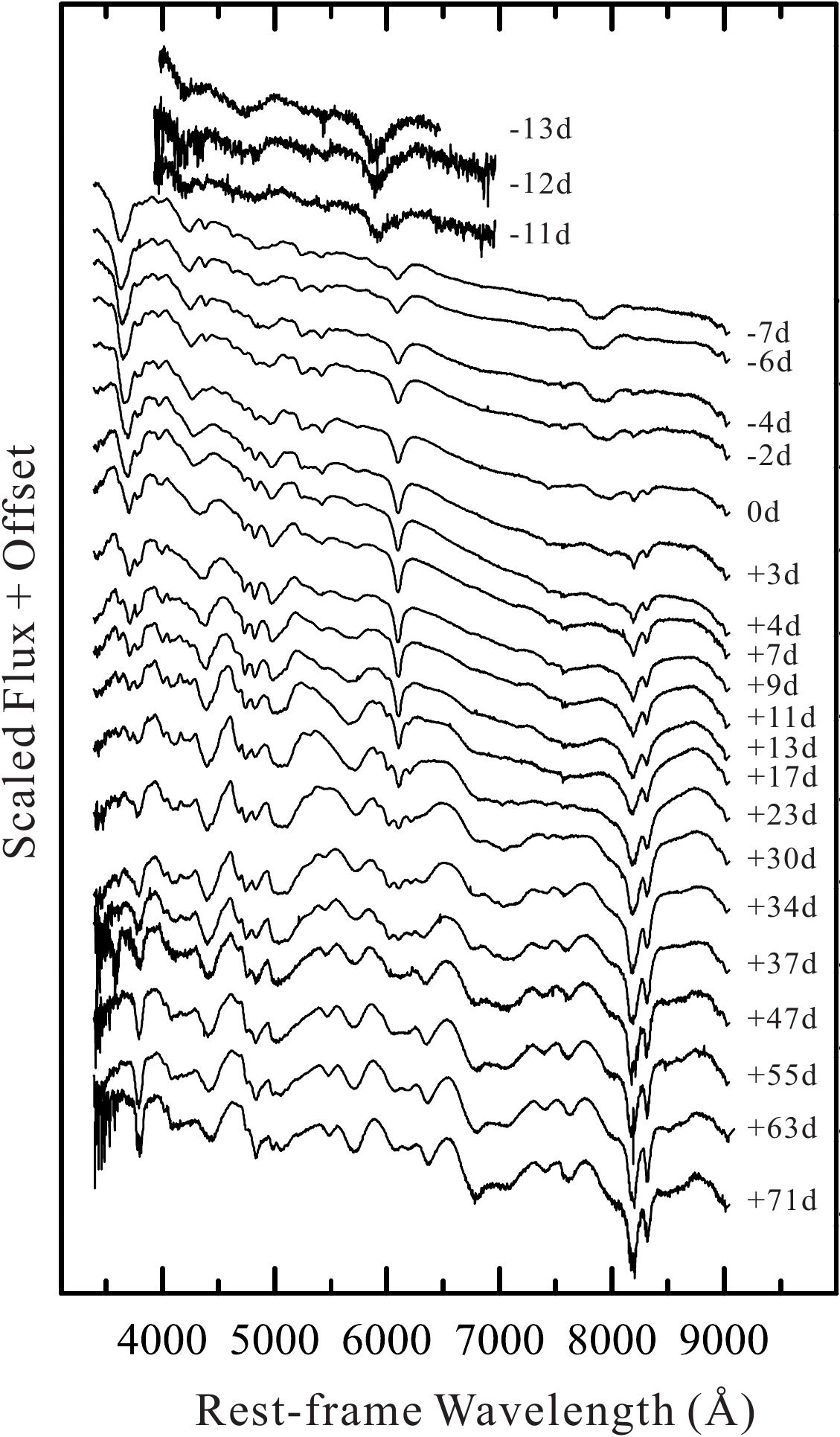}
\caption{Optical spectral evolution of SN 2012fr. The spectra have been corrected for the redshift of the host galaxy ($V_{hel}$ = 1636 km s$^{-1}$) and telluric lines. They have been shifted vertically by arbitrary amounts for clarity; the numbers on the right-hand side mark the epochs of the spectra in days after $B$ maximum.}
\label{<Spe_whole>}
\end{figure}

A total of 20 spectra of SN 2012fr were obtained with the LJT and YFOSC spectrograph, which covers a phase from $t\approx -7$ to $t\approx +71$ days relative to the $B$-band maximum light. The complete spectral evolution of SN 2012fr is presented in Figure \ref{<Spe_whole>}, where the three earliest spectra obtained through ARAS are also overplotted. The early-time ARAS spectra show strong absorption features at $5900\sim6000$ \AA~, perhaps due to the largely blueshifted \ion{Si}{2} $\lambda$6355. At around the maximum brightness, the spectral evolution of SN 2012fr generally follows that of normal SNe Ia but showing narrower and weaker absorption features of \ion{Si}{2} 6355. The detailed spectral evolution is discussed in the following subsections.

\subsection{Spectra Temporal Evolution}
\label{subsect:Temsp}
\begin{figure}[!th]
\centering
\includegraphics[width=5cm, angle=0]{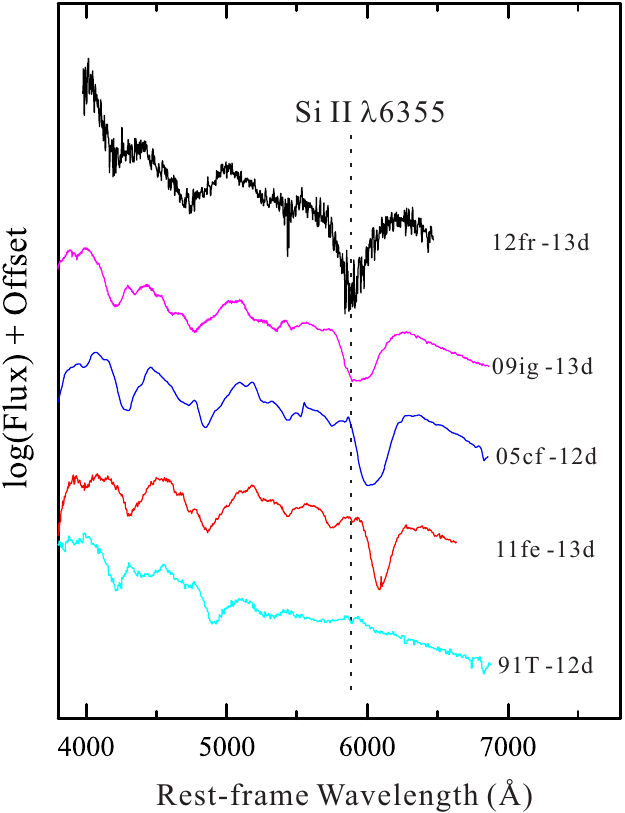}
\caption{Spectrum of SN 2012fr at $t = -13$ days from $B$-band maximum, overplotted with the spectra of SN 1991T, SN 2005cf, SN 2009ig, and SN 2011fe at similar phases (see text for references). All spectra presented here have been corrected for redshift and reddening. The dashed line marks the absorption minimum of \ion{Si}{2} $\lambda$ 6355 for SN 2012fr. }
\label{<Sp_ear_com>}
\end{figure}

\begin{figure*}[!th]
\centering
\includegraphics[width=15.5cm, angle=0]{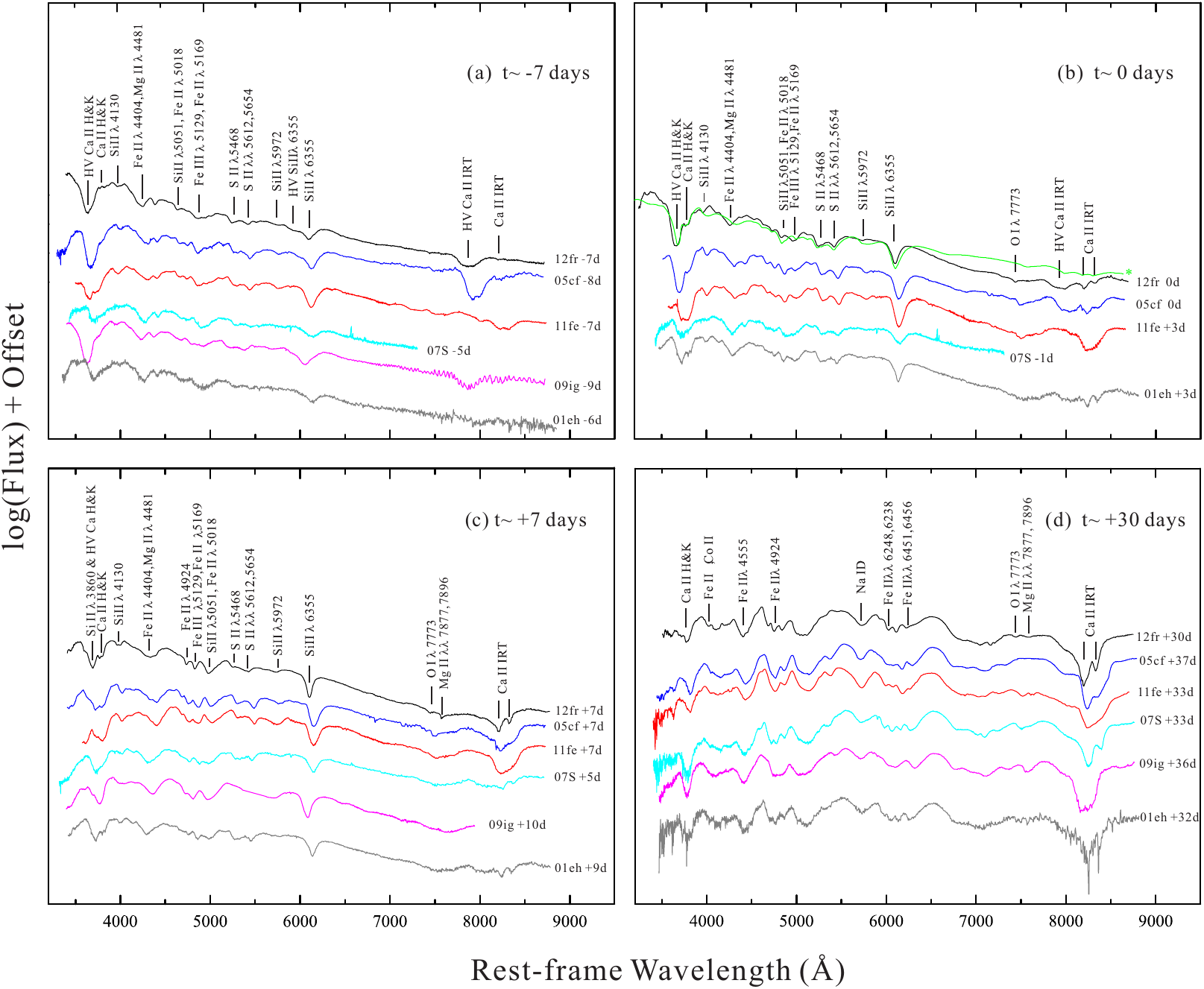}
\caption{Spectra of SN 2012fr at t $\approx$ $-$7, 0, +7 and +30 days after the $B$-band maximum. Overplotted are the comparable-phase spectra
of SN 2001eh,  SN 2005cf, SN 2007S, SN 2009ig, and SN 2011fe (see text for references). The green line in panel (b) is the SYNOW fit
to the t$\approx$0 day spectrum of SN 2012fr.}
\label{<4epoch>}
\end{figure*}

The spectral comparisons between SN 2012fr and other well-observed SNe Ia are shown in Figures \ref{<Sp_ear_com>} and \ref{<4epoch>}. All the spectra have been corrected for redshift but not for the reddening.

Figure \ref{<Sp_ear_com>} displays the spectrum of SN 2012fr taken at $t=-13$ days, with typical features of $W$-shaped sulfur lines and iron lines at 4500$\sim$5000\AA. The most notable feature is the strong absorption trough at $\sim$ 5900\AA~, which can be attributed to the \ion{Si}{2} $\lambda$6355 absorption formed in regions above the photosphere. Such a HVF is similarly seen in SN 2009ig and SN 2005cf, although SN 2005cf shows a relatively lower velocity. Note that this detached HVF is not observed in the spectrum of SN 2011fe, suggesting that it may be not a common feature for the spectroscopically normal SNe Ia. As it can be seen from the plot, the \ion{Si}{2} feature is barely detected in SN 1991T at such an early phase (Mazzali et al. 1995). In addition, we notice that the \ion{C}{2} 6580 absorption can be detected in SN 2005cf and SN 2011fe, but not in SN 1991T and SN 2009ig. It is not clear whether SN 2012fr shows the \ion{C}{2} $\lambda$6580 absorption at this phase because of a relatively poor S/N ratio for the spectrum (but see discussions below).

Figure \ref{<4epoch>} shows the spectroscopic comparison of SN 2012fr with some SNe Ia at several epochs. The comparison sample was selected because they are well-observed or show similarities to SN 2012fr in some respects, and includes SN 2001eh (Silverman et al. 2012), SN 2005cf (Wang et al. 2009b), SN 2007S (Blondin et al. 2012), SN 2009ig (Foley et al. 2012), and SN 2011fe (Pereira et al. 2013; Wang et al. 2014 in preparations). At one week before the maximum, the spectrum is characterized by lines of singly ionized intermediate-mass elements (IMEs, e.g., Si, S, Mg and Ca). The absorption feature at 6000\AA~ is dominated by the photospheric component of \ion{Si}{2} $\lambda$6355; and the HVF becomes nearly invisible in all of our sample. There is no significant signature of C II 6580 absorption in the t = $-$7 day spectrum of SN 2012fr, consistent with the conclusion reached by C13 from their earlier spectra. We noticed that such a C II feature is not visible in SN 2001eh, SN 2007S, and 2009ig. By $t = -7 $ days, the HVFs of \ion{Ca}{2} IR triplet are still prominent and are stronger than the photospheric components in SN 2012fr and SN 2005cf, while they are barely visible in SN 2001eh and SN 2011fe. This indicates that the distribution of \ion{Ca}{2} in the outer layer of the ejecta may provide another significant signature to identify the diversity of SN Ia explosions. A minor absorption at $\sim$ 5750 \AA~ is likely due to \ion{Si}{2} 5972, which is absent in the early spectra; the weak strength suggests a relatively high temperature for the photosphere with respect to normal SNe Ia (Nugent et al. 1997).

Around maximum brightness, the absorption feature of \ion{Si}{2} $\lambda$6355 in SN 2012fr evolves as a relatively normal profile. Nevertheless, the line profile of \ion{Si}{2} $\lambda$6355 appears apparently narrower compared to those of the normal SNe Ia like SN 2005cf and SN 2011fe. The HVFs of \ion{Ca}{2} IR triplet absorption feature are comparably strong in SN 2012fr and SN 2005cf; but they are nearly invisible seen in SN 2011fe where the absorption trough is dominated by the photospheric component. The \ion{O}{1} $\lambda$7773 feature strengthens in all cases for our sample, while it is not clear for SN 2007S because of the limited wavelength coverage. The line-strength ratio of \ion{Si}{2} 5972 to \ion{Si}{2} 6355, known as $R$(\ion{Si}{2}) (Nugent et al. 1997), is an approximate indicator of the photospheric temperature, with a larger value corresponding to a lower temperature. We measured this parameter to be 0.07$\pm$0.02 for SN 2012fr near the maximum light, which is noticeably smaller than the corresponding value for the standard Ia SN 2005cf, suggesting a higher photospheric temperature for SN 2012fr.

At $t \sim +7$ days, our sample generally exhibit a similar spectral evolution, with the HVFs becoming undetectable in the main spectral lines. One interesting feature is that the absorption trough of \ion{Ca}{2} IR feature clearly splits into two components in SN 2012fr, SN 2001eh, and SN 2007S. This behavior is also consistent with the narrow \ion{Si}{2} profiles seen in these three events, suggesting that the ejecta produced from their explosions may be confined into a smaller range of velocity compared to SN 2005cf, SN 2009ig, and SN 2011fe. At $t\sim1$ month (Figure \ref{<4epoch>}d), the spectra are dominated by iron lines and a strong \ion{Ca}{2} IR triplet absorption trough. The double absorption features of \ion{Ca}{2} IR triplet become more prominent in SN 2012fr, SN 2001eh, and SN 2007S relative to the rest comparison SNe Ia.

The above comparison suggests that SN 2012fr shares similar features with SN 2001eh and SN 2007S in the near-maximum and post-maximum phases; and it appears to be more similar to the high-velocity SN Ia 2009ig in the very early phase. To examine the mechanism forming the narrower \ion{Si}{2} and \ion{Ca}{2} lines, we adopted the parameterized resonance scattering synthetic-spectrum code SYNOW (Fisher et al. 1999; Branch et al. 2005) to fit the t = 0 day spectrum of SN 2012fr (see green line in Figure 8b). The radial dependence of the line optical depth is chosen to be exponential with an e-folding velocity v$_{e}$ (i.e., $\tau \propto$ exp($-$v/v$_{e}$)). We found that the v$_{e}$ should be smaller than 1000 km s$^{-1}$ in order to fit the narrow \ion{Si}{2} absorption at 6100\AA~ in SN 2012fr, suggesting that the silicon layer (and perhaps the \ion{Ca}{2} layer) becomes optically thin near the photosphere.

\subsection{The Ejecta Velocity}

In this subsection, we examine the ejecta velocity of SN 2012fr via the absorption features of some spectral lines. The location of the blueshifted absorption minimum was measured by using both the Gaussian fit routine and the direct measurement of the center of the absorption, and the results were averaged.
\begin{figure}[!th]
\centering
\includegraphics[width=8.5cm,angle=0]{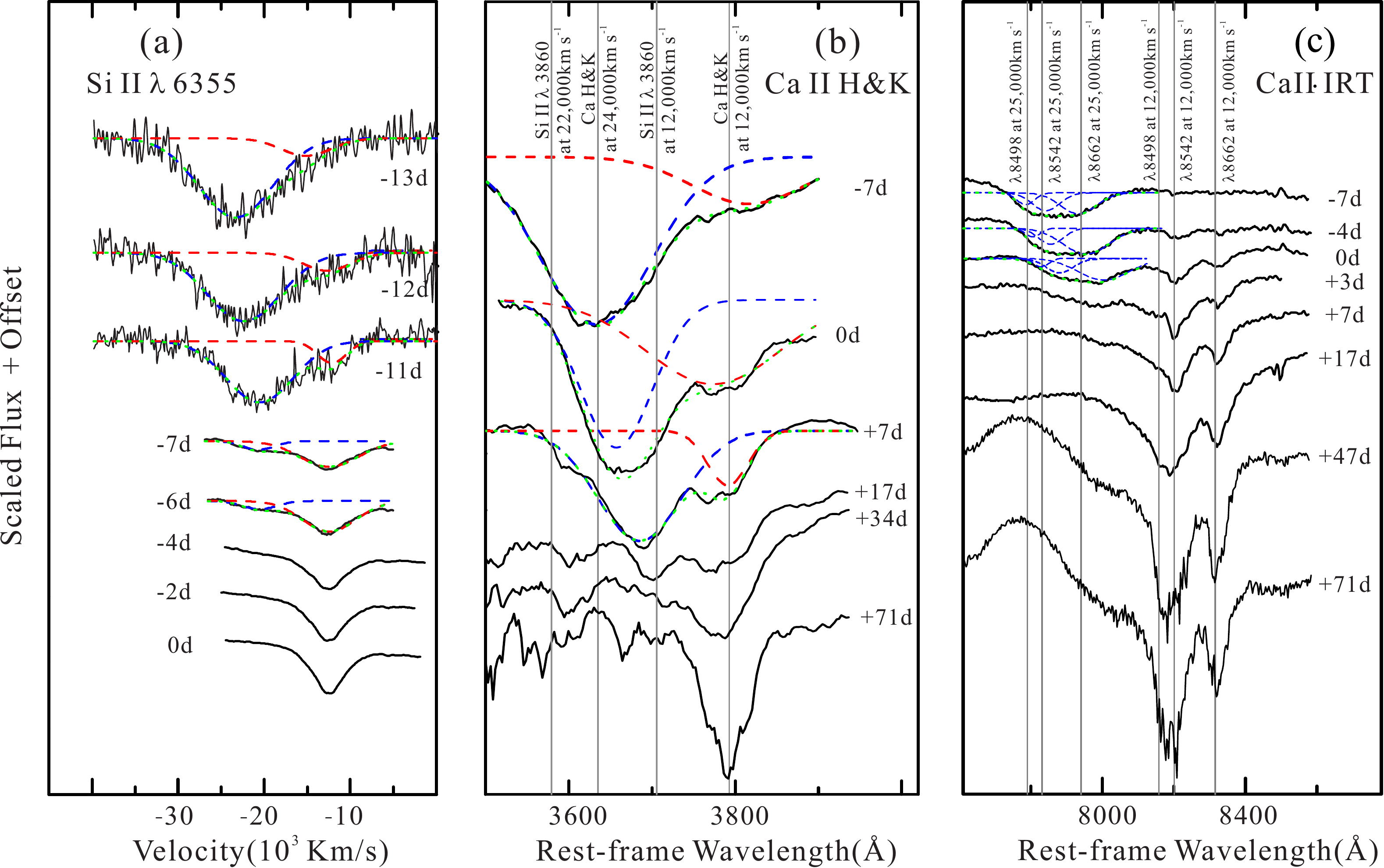}
\caption{The evolution of HVFs of \ion{Si}{2} and \ion{Ca}{2} in the spectra of SN 2012fr. Panel (a): \ion{Si}{2} $\lambda$6355; Panel (b): \ion{Ca}{2} H\&K; Panel (c): \ion{Ca}{2} IR triplet. In panels (a) and (b), the blue and red dashed lines represent the Gaussian fit to the detached high-velocity and the photospheric components, respectively; green dotted line denotes the sum of these two components. The blue dashed lines in panel (c) represent the triple-Gaussian fit to the HVFs of \ion{Ca}{2} IR absorption, and the green dotted line is the sum of the Gaussian fit. The wavelength positions of the Ca II H\&K and Ca II IR lines at different velocities are marked as a guide to the eyes.}
\label{<MPF_HVFs>}
\end{figure}

\begin{figure}[!th]
\centering
\includegraphics[width=8.5cm,angle=0]{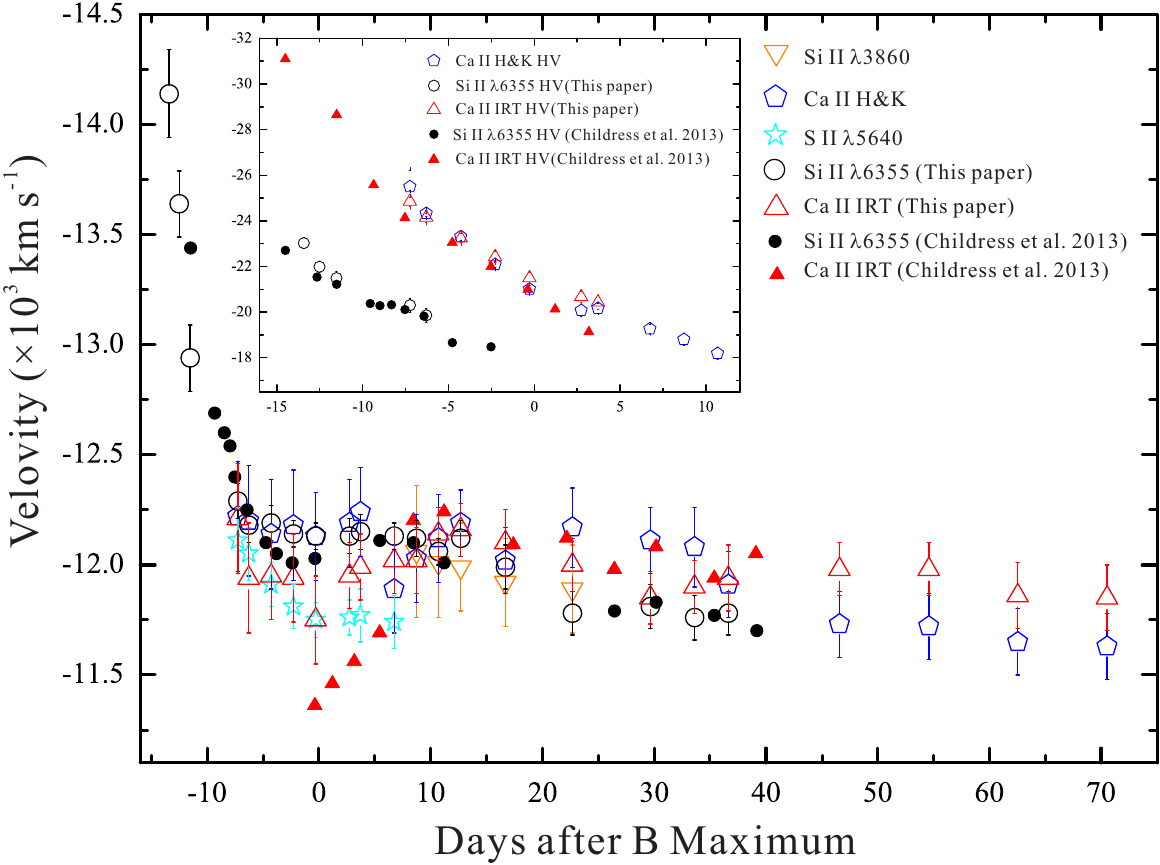}
\caption{Velocity evolution of different elements inferred from the spectra of SN 2012fr. The inset plot shows the velocity evolution of the detached, high-velocity components. All of the velocity data are listed in Table \ref{Tab:Velo_Ev}.}
\label{<Vabs_all>}
\end{figure}

Figure \ref{<MPF_HVFs>} shows the evolution of the line profiles of \ion{Si}{2} $\lambda$6355, \ion{Ca}{2} IR triplet, and \ion{Ca}{2} H$\&$K. For these lines, the blue-side absorption feature gradually weakens with the emergence of the red-side component; such an overall evolution clearly suggests the presence of another HV component in these spectral features. Adopting the double-Gaussian fit and attributing the absorption on the blue side to the high-velocity component, we estimate from the t = $-$7 day spectrum that the detached components of \ion{Si}{2} $\lambda$6355 and \ion{Ca}{2} H\&K have a velocity of $\sim$22,000 km s$^{-1}$ and $\sim$25,000 km s$^{-1}$, respectively. These velocities are much higher than the corresponding photospheric velocities (e.g., $v_{phot}$$\sim$12,000 km s$^{-1}$; see Figure \ref{<Vabs_all>}).

Inspection of Figure \ref{<MPF_HVFs>}(a) reveals that the detached high-velocity component of the \ion{Si}{2} line almost disappears in the t = $-$7 days spectrum, indicating that the HVF of \ion{Si}{2} line is detectable for only a very short time in some SNe Ia. In contrast, the corresponding HVFs of the \ion{Ca}{2} IR triplet can last for a longer time and is still detectable in the t = + 7 days spectrum of SN 2012fr. The duration for the HVFs of \ion{Ca}{2} H\&K lines, as shown is Figure \ref{<MPF_HVFs>}(b), is unclear due to the possible contamination of the \ion{Si}{2} $\lambda$3860 line. This needs to be clarified, as the \ion{Ca}{2} H\&K absorption feature may be a potential indicator of SN Ia diversity (e.g., Maguire et~al. 2012). As shown in Figure \ref{<MPF_HVFs>}(b), the fits using detached high-velocity and photospheric Ca II match well with the observed spectra; the HVF slows down with time and the photospheric component remains at a constant velocity of 12,000 km s$^{-1}$ (see also Fig. 10). In contrast, neither the high-velocity nor the photospheric component of \ion{Si}{2}$\lambda$3860 (see the dashed lines in the plot) match the two absorption features of Ca II H\&K lines before t = +7 days. Thus there does not seem to be any need for \ion{Si}{2} 3860 in those profiles of earlier spectra (but see \citealp{foley13}). However, the \ion{Si}{2} $\lambda$3860 may contribute to the \ion{Ca}{2} H\&K absorption trough in the later spectra. For example, the absorption near 3720\AA~ in the $t\approx +17$ spectrum might be due to the photospheric \ion{Si}{2} $\lambda$3860 line (see also Table \ref{Tab:Velo_Ev}), as the HVFs of Ca II should have disappeared at this late phase (see also \ref{<MPF_HVFs>}(c)).

Figure \ref{<MPF_HVFs>}(c) shows the evolution of \ion{Ca}{2} IR triplet absorption. The HVFs dominate the line profiles in the earlier phase (see also C13 ), but the substructures of \ion{Ca}{2} $\lambda$8498, $\lambda$8542\AA, and $\lambda$8662 absorptions are hardly separated because of the blending. In the $t = -4$ days spectrum, the photospheric components are visible and the weak absorption features near 8200\AA~ and 8350\AA~ can be identified as \ion{Ca}{2} 8542 (at $\sim$ 12,100 km s$^{-1}$) and \ion{Ca}{2} 8662 (at $\sim$ 12,100 km s$^{-1}$), respectively. The \ion{Ca}{2} $\lambda$8498 absorption is marginally detected at $\sim$8160\AA~ in the t = +3 day spectrum, but it seems to become progressively strong with time along with  \ion{Ca}{2} 8542 and \ion{Ca}{2} 8662 in the later spectra. It is worth mentioning that the three components of \ion{Ca}{2} IR triplet absorptions show a clear separation in the t = +71 day spectrum. The narrow line wings indicate that the ejecta is perhaps confined into a narrow range of velocity.

\begin{deluxetable*}{ccccccccc}[!th]

\tablewidth{0pt}

\tabletypesize{\scriptsize}
\tablecaption{Velocity Evolution of Different Lines from SN 2012fr}
\tablehead{\colhead{Epoch} & \colhead{Si II} & \colhead{HVFs} & \colhead{Ca II } & \colhead{S II} & \colhead{HVFs} & \colhead{Si II} & \colhead{HVFs} & \colhead{Ca II} \\
\colhead{(days)} & \colhead{($\lambda$3860)} & \colhead{(H$\&$K)} & \colhead{(H$\&$K)} & \colhead{($\lambda$5640)} & \colhead{($\lambda$6355)} & \colhead{($\lambda$6355)} & \colhead{(IRT)} & \colhead{(IRT)} }

\startdata
-13.43	&	\nodata	&	\nodata	&	\nodata	&	\nodata	&	-23.15(25)\tablenotemark{a}	&	-15.14(20)\tablenotemark{a}	&	\nodata	&	\nodata	\\
-12.51	&	\nodata	&	\nodata	&	\nodata	&	\nodata	&	-22.11(25)\tablenotemark{a}	&	-13.64(15)\tablenotemark{a}	&	\nodata	&	\nodata	\\
-11.53	&	\nodata	&	\nodata	&	\nodata	&	\nodata	&	-21.61(30)\tablenotemark{a}	&	-12.94(15)\tablenotemark{a}	&	\nodata	&	\nodata	\\
-7.25	&	\nodata	&	-25.50(15)	&	-12.22(25)	&	-12.11(15)	&	-20.41(30)\tablenotemark{a}	&	-12.29(08)	&	-12.01(25)	&	-24.98(20)	\\
-6.29	&	\nodata	&	-24.33(15)	&	-12.20(25)	&	-12.05(12)	&	-19.97(30)\tablenotemark{a}	&	-12.18(08)	&	-11.94(25)	&	-24.28(20)	\\
-4.28	&	\nodata	&	-23.31(15)	&	-12.14(25)	&	-11.91(10)	&	\nodata	&	-12.19(08)	&	-11.95(20)	&	-23.37(20)	\\
-2.28	&	\nodata	&	-22.09(15)	&	-12.18(25)	&	-11.81(10)	&	\nodata	&	-12.14(06)	&	-11.94(20)	&	-22.54(25)	\\
-0.27	&	\nodata	&	-21.01(15)	&	-12.13(20)	&	-11.75(08)	&	\nodata	&	-12.13(06)	&	-11.94(20)	&	-21.61(25)	\\
2.74	&	\nodata	&	-20.07(25)	&	-12.19(20)	&	-11.76(08)	&	\nodata	&	-12.13(08)	&	-11.95(15)	&	-20.78(25)	\\
3.72	&	\nodata	&	-20.16(20)	&	-12.24(20)	&	-11.77(12)	&	\nodata	&	-12.15(08)	&	-11.99(15)	&	-20.54(30)	\\
6.74	&	\nodata	&	-19.27(20)	&	-11.89(20)	&	-11.74(12)	&	\nodata	&	-12.13(06)	&	-12.02(15)	&	\nodata	\\
8.72	&	-12.06(30)	&	-18.80(25)	&	-12.03(20)	&	\nodata	&	\nodata	&	-12.12(08)	&	-12.02(15)	&	\nodata	\\
10.68	&	-12.01(25)	&	-18.19(25)	&	-12.12(20)	&	\nodata	&	\nodata	&	-12.06(06)	&	-12.14(12)	&	\nodata	\\
12.7	&	-11.99(20)	&	\nodata	&	-12.19(15)	&	\nodata	&	\nodata	&	-12.12(08)	&	-12.16(12)	&	\nodata	\\
16.67	&	-11.92(20)	&	\nodata	&	-12.02(15)	&	\nodata	&	\nodata	&	-11.99(10)	&	-12.10(15)	&	\nodata	\\
22.67	&	-11.89(20)	&	\nodata	&	-12.17(18)	&	\nodata	&	\nodata	&	-11.78(10)	&	-12.00(12)	&	\nodata	\\
29.64	&	\nodata	&	\nodata	&	-12.11(15)	&	\nodata	&	\nodata	&	-11.81(10)	&	-11.85(12)	&	\nodata	\\
33.61	&	\nodata	&	\nodata	&	-12.08(18)	&	\nodata	&	\nodata	&	-11.76(10)	&	-11.90(12)	&	\nodata	\\
36.63	&	\nodata	&	\nodata	&	-11.91(15)	&	\nodata	&	\nodata	&	-11.78(10)	&	-11.94(15)	&	\nodata	\\
46.6	&	\nodata	&	\nodata	&	-11.73(15)	&	\nodata	&	\nodata	&	\nodata	&	-11.98(12)	&	\nodata	\\
54.6	&	\nodata	&	\nodata	&	-11.72(15)	&	\nodata	&	\nodata	&	\nodata	&	-11.98(12)	&	\nodata	\\
62.55	&	\nodata	&	\nodata	&	-11.65(15)	&	\nodata	&	\nodata	&	\nodata	&	-11.86(15)	&	\nodata	\\
70.54	&	\nodata	&	\nodata	&	-11.63(15)	&	\nodata	&	\nodata	&	\nodata	&	-11.85(15)	&	\nodata	\\
\enddata

\tablecomments{Uncertainties, in units of 10 km s$^{-1}$, are 1 $\sigma$.}
\tablenotetext{a}{Deduced from double-Gaussian fit.}

\label{Tab:Velo_Ev}
\end{deluxetable*}

The derived velocities of the intermediate-mass elements (IMEs) such as S, Si, and Ca as a function of time are shown in Figure \ref{<Vabs_all>} and Table \ref{Tab:Velo_Ev}. Overplotted are the corresponding velocities obtained by C13. One can see that our results agree with theirs to within 100 km s$^{-1}$ at similar phases. We also notice that the photospheric velocities of \ion{Ca}{2} IR triplet derived in C13 are slower than ours at t $<$ +6 days. This difference might be due to the uncertainties in the fitting process of the line profiles, as the photospheric components of \ion{Ca}{2} IR triplet are relatively weak in the earlier phases and may not be appropriately indicated with the Gaussian profiles. In SN Ia spectra, the \ion{Si}{2} $\lambda$6355 line can better trace the velocity of the photosphere because this feature suffers less blending with other lines. After the maximum light, the velocity of \ion{Si}{2} 6355 remains at $\sim$12,000 km s$^{-1}$ for over a month, with a velocity gradient of $2\pm13$ km s$^{-1}$ day$^{-1}$. Such a low velocity gradient is consistent with that derived from C13 at similar phases (e.g., 0.3$\pm$10 km s$^{-1}$ day$^{-1}$). Such a low velocity gradient is similarly seen for the velocity evolution of \ion{Ca}{2}, which clearly puts SN 2012fr into the LVG category of SNe Ia according to the classification scheme of Benetti et al (2005). As it can be seen from Figure \ref{<Vabs_all>}, the IMEs of SN 2012fr generally have similar expansion velocities (e.g., $\sim$12,000 km s$^{-1}$), suggestive of a relatively uniform distribution of the burning products in the ejecta. Table \ref{table:Par_SN} collects the basic information for SN 2012fr and its host galaxy.

\section{Discussions}
\label{sect:Discu}

\subsection{The Peak Luminosity and the Nickel Mass}
To further examine the properties of SN 2012fr, it is important to know the luminosity of SN 2012fr and the nickel mass produced in the explosion. The distance to SN 2012fr is important for deriving these two quantities. Direct measurements to its host galaxy NGC 1365 are available through several methods. In our analysis, we adopt the most recent measurement from the Tully-Fisher relation which gives a distance modulus $\mu$ = 31.38 $\pm$ 0.06 mag (Tully et al. 2009) or a distance of 18.9 Mpc. Adopting this distance and correcting for the galactic extinction, we derive the absolute $B$- and $V$-band peak magnitudes as M$_B = -19.49 \pm$ 0.06 mag and M$_V = -19.48 \pm$ 0.06 mag, respectively. This indicates that SN 2012fr is brighter than a typical SN Ia (e.g., M$_{V}$ = $-$19.27 mag from Wang et al. 2006) by $\sim$20\%.

\begin{deluxetable}{cc}[!th]
\tablewidth{0pt}
\tabletypesize{\small}
\tablecaption{Relevant Parameters for NGC 1365 and SN 2012fr}
\tablehead{\colhead{Parameter } & \colhead{Value}}
  \startdata
 \cutinhead{ Parameters for NGC 1365}
 Galaxy Morphology & SBb  \\
 Activity type & Seyfert 1.8\\
m-M  & 31.38 $\pm$ 0.06 mag\tablenotemark{a} \\
$E(\bv)_{Gal}$ & 0.02\tablenotemark{b}\\
V$_{hel}$  & 1636 $\pm$ 1 km s$^{-1}$ \\
 \cutinhead{Parameters for SN 2012fr}
Discovery Date & Oct. 27 2012 \\
Epoch of B Maximum & 2456243.50(Nov. 12 2012) \\
m$_B$  &  12.01 $\pm$ 0.03 mag \\
m$_V$  &  11.99 $\pm$ 0.03 mag \\
$\Delta$m$_{15}(B)$  & 0.85 $\pm$ 0.05 mag   \\
M$_{B}$  & -19.49 $\pm$ 0.06 mag \\
M$_{V}$  & -19.48 $\pm$ 0.06 mag \\

L$^{max}_{bol}$ & (1.82$\pm$ 0.15) $\times 10^{43}$ erg s$^{-1}$\\
$^{56}$Ni & 0.88 $\pm$ 0.08 M$_{\sun}$ \\
$v_{max}$ & 12,120 $\pm$ 70  km s$^{-1}$  \\
$v_{10}$  & 12,100 $\pm$ 100  km s$^{-1}$   \\
$\dot{v}_{10}$ & 2 $\pm$  13 km s$^{-1}$day$^{-1}$ \\
$R$(Si) & 0.07 $\pm$ 0.02 \\

\enddata
\tablecomments{$v_{max}$ is the velocity of \ion{Si}{2} $\lambda$6355 measured at the $B$-band maximum; $v_{10}$(\ion{Si}{2}) represents the
value obtained at t = +10 days. The velocity gradient during this period is defined as $\dot{v}_{10}$ (Benetti et al. 2005); $R$(Si) represents the ratio of \ion{Si}{2} 5972 and \ion{Si}{2} 6355 lines at the $B$-band maximum. }
\tablenotetext{a}{Measured by Tully-Fisher relation (Tully et al. 2009).}
\tablenotetext{b}{\citet{Schle98}}
\label{table:Par_SN}
\end{deluxetable}

\label{subsect:Bolmag}
The bolometric luminosity of SN 2012fr is derived from the UV and optical photometric observations. The near-infrared (NIR) emission is corrected according to the flux ratio derived by Wang et al. (2009b). They found that the ratio of the NIR-band emission (9000--24,000~\AA) to the optical (3200--9000\AA) lies between 5-20\% for normal SNe~Ia like SN 2005cf, depending on the supernova phase. An uncertainty of 10\% is assumed for the NIR corrections in the derivation of bolometric luminosity. Considering the SN Ia emission as a blackbody radiation, the missing UV flux at wavelengths shorter than $Swift$ UV filters (e.g., $<$ 1600 \AA) is estimated to be $<$ 3.0\% based on the spectral template of Hsiao et al. (2007), we thus ignored its contribution in the calculation.

The peak bolometric luminosity of SN 2012fr is estimated to be (1.82 $\pm$ 0.15) $\times$ 10$^{43}$ erg s$^{-1}$. The uncertainty includes the errors in the distance modulus, the observed magnitudes, and the NIR corrections. With the derived bolometric luminosity, the synthesized $^{56}$Ni mass can be estimated using the Arnett law (Arnett 1982; Stritzinger \& Leibundgut 2005):
\begin{equation}
L_{max}=(6.45e^{\frac{-t_r}{8.8d}}+1.45e^{\frac{-t_r}{111.3d}})(\frac{M_{Ni}}{M_{\sun}})\times10^{43} erg~s^{-1},
\end{equation}
where $t_r$ is the rise time of the bolometric light curve, and $M_{Ni}$ is the $^{56}$Ni mass (in units of solar masses, M$_{\sun}$). From the discovery information \citep{CEBT3277} of SN 2012fr, we estimate the rise time to be about 18 days in the $V$ band. Inserting this value and the maximum bolometric luminosity into Equation (1), we derive a nickel mass of 0.88 $\pm$ 0.08M$_{\sun}$ for SN 2012fr. This value is smaller than that of SN 1991T (e.g., 1.1M$_{\sun}$; Contardo et al. 2006) and larger than that of SN 2005cf (e.g., 0.78 M$_{\sun}$; Wang et al. 2009b), SN 2009ig (e.g., 0.80 M$_{\sun}$; Sahu et al. 2011), and SN 2011fe (e.g., 0.53 M$_{\sun}$; Pereira et al. 2013).

\begin{figure*}[!th]
\centering
\includegraphics[height=11cm, angle=0]{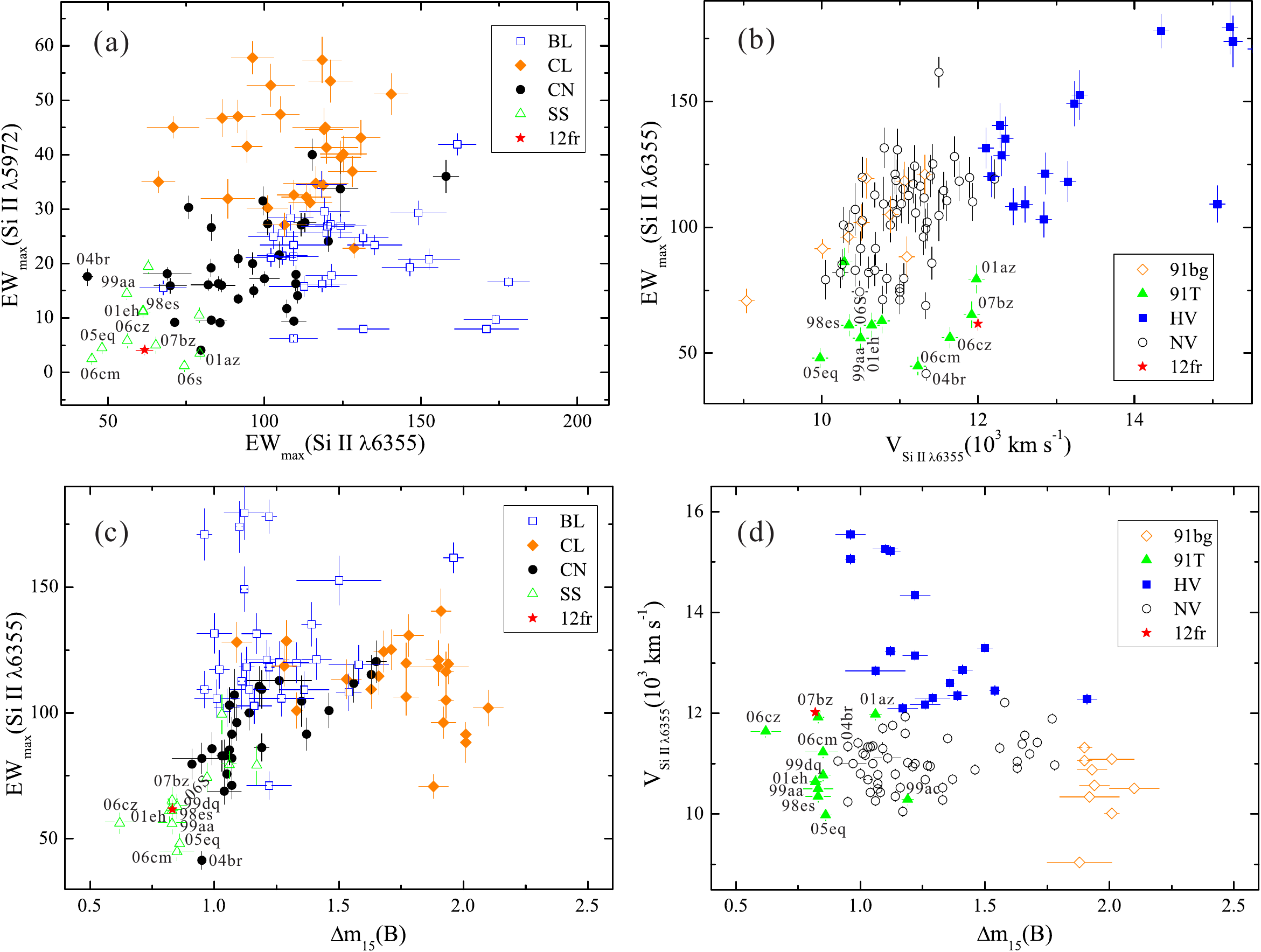}
\caption{Comparison of various spectroscopic indicators from SN 2012fr with those from other SNe Ia as measured by Blondin et~al. (2012), Silverman et~al. (2012), Wang et al. (2009a), and this paper. The selected sample have spectra within $\pm3$ days from $B$-band maximum. Panel (a):  the \ion{Si}{2} $\lambda$5972 vs. \ion{Si}{2} $\lambda$6355 at maximum light with spectroscopic subclasses defined by Branch et al. (2009);  panel (b): the $EW$ vs. velocity of \ion{Si}{2} 6355 at maximum light with subclasses defined by Wang et al. (2009a); Panels (c) and (d): the $EW$ and $V_\mathrm{Si~II \lambda 6355}$  vs. $\Delta m _{15}$(B) at maximum light.}
\label{<Diversity>}
\end{figure*}

\subsection{The Spectroscopic Classification}
\label{sect:Diversity}
\citet{Blondin12} and \citet{Silver12} presented detailed comparison study of different classification schemes based on the large SNe Ia spectral datasets from the CfA and Berkeley SuperNova Ia Program, respectively. From the near-maximum light spectrum of SN 2012fr, the equivalent width of \ion{Si}{2} $\lambda$6355 and \ion{Si}{2} $\lambda$5972 lines is measured to be about 60\AA ~ and  4\AA ~ (see Table 8), respectively. The weak \ion{Si}{2} absorption shows some similarities to the SS/91T subclass of SNe Ia, though its location is also close to the core-normal subclass in the Branch diagram, as shown in Figure \ref{<Diversity>}a and c. On the other hand, SN 2012fr can be clearly put into the LVG group of the classification system of Benetti et al. (2005) in terms of the lower velocity gradient. In the Wang et al. diagram, however, SN 2012fr resides between the HV and the 91T-like subclasses (Figure \ref{<Diversity>}b and d). SN 2012fr has an expansion velocity slightly higher than a normal SN Ia, but it bears many properties that are not seen in the HV subclass. The weak \ion{Si}{2} absorption (see Table \ref{Tab:pEW}) near the maximum light, the lower velocity gradient, and the higher luminosity are all reminiscent of the properties of the 91T-like objects. Although the HVFs of \ion{Ca}{2} IR are also found to be relatively strong for those events \citep{ChildHVF}, the absolute strength of these HVFs are usually very weak in the 91T-like objects (e.g., with EW $<$ 50\AA~). Moreover, SN 2012fr has very strong HVF of \ion{Si}{2}$\lambda$6355 but this feature is rarely detected in the 91T-like events. This discrepancy further complicates the classification of SN 2012fr.

\begin{deluxetable}{cccc}[!th]
\tabletypesize{\small}
\tablewidth{0pt}
\tablecaption{The Pseudo EW of \ion{Si}{2} Absorption Line of SN 2012fr}
\centering

\tablehead{\colhead{Epoch} & \colhead{$\lambda$5972} & \colhead{$\lambda$6355} & \colhead{HVFs($\lambda$6355)}  }
\startdata
-13.43	&	\nodata	&	21.6	&	103.8	\\
-12.51	&	\nodata	&	26.0	&	138.1	\\
-11.53	&	\nodata	&	32.2	&	98.6	\\
-7.25	&	4.8	&	50.5	&	20.86	\\
-6.29	&	7.2\tablenotemark{a}	&	51.6	&	18.8	\\
-4.28	&	7.7\tablenotemark{a}	&	57.9	&	\nodata	\\
-2.28	&	4.3	&	62.0	&	\nodata	\\
-0.27	&	4.1	&	61.8	&	\nodata	\\
+2.74	&	9.1\tablenotemark{a}	&	67.8	&	\nodata	\\
+3.72	&	10.1\tablenotemark{a}	&	62.4	&	\nodata	\\
+6.74	&	7.9	&	62.2	&	\nodata	\\
+8.72	&	6.2	&	65.8	&	\nodata	\\
+10.68	&	\nodata	&	62.5	&	\nodata\\	
+12.70	&	\nodata	&	60.5	&\nodata	\\	
+16.67	&	\nodata	&	65.5	&	\nodata	\\
\enddata

\tablecomments{The pEW of \ion{Si}{2} $\lambda$5972, $\lambda$6355 absorption in the spectra of SN 2012fr, in units of \AA.}
\tablenotetext{a}{Blended with bluer absorption line. }

\label{Tab:pEW}
\end{deluxetable}

To examine whether SN 2012fr forms a separate subgroup of SNe Ia, we attempt to collect a sample like SN 2012fr to get a better statistical study of their photometric and spectroscopic behaviors. Figure \ref{<Diversity>} shows the distribution of SNe~Ia\footnote{The spectral and photometric data used in the analysis are taken from Blondin et al. 2012, Silverman et al. 2012, Hicken et al. 2009, and Ganeshalingam et al. 2010}. We found that SN 2012fr and a few other SNe~Ia (including SNe 2001eh, 2004br, 2005eq, 2006cm, and 2007bz) cluster tightly together in the plot, all showing unusually narrow line profiles of \ion{Si}{2}$\lambda$ 6355 and Ca II IR triplet. Another common feature for these objects is that they all have smaller light-curve decline rates, with $\Delta m_{15}$ $\lesssim$ 1.0 mag. It is notable that most of the Narrow-Lined (NL) SNe Ia can be also put into the SS/91T-like subclass. Inspection of Figure \ref{<Diversity>}d reveals that the NL subclass also overlap largely with the 91T-like subclass, with 10, 000 km s$^{-1}$$\lesssim v_\mathrm{Si~II}$ $\lesssim$ 12,000 km s$^{-1}$.

The above analysis indicates that SN 2012fr could be a subset of the SS/91T-like subclass. The differences seen in the earlier phases, such as the HVFs of \ion{Si}{2} and \ion{Ca}{2}, may be explained by the viewing-angle effect. Maund et~al. (2013) obtained the spectropolarimetric measurements of SN 2012fr and found the high-velocity components are highly polarized in the early phases, with the degree of polarization being 0.40\% for Si II line and 0.85\% for Ca II lines. At around the maximum light, the polarizations for the photospheric components of Si II and Ca II are found to be 0.65\% and 0.54\%, respectively. On the other hand, the line polarizations of 91T-like SNe Ia are usually at low levels (e.g., $<$ 0.20\%) according to Wang et al. (2007). We therefore propose that SN 2012fr may be a counterpart of the 91T-like SNe Ia, but it was viewed in a direction where the ejecta has a clumpy or shell-like structure.

\section{Conclusion}
\label{sect:con}
We have presented the ultraviolet and optical observations of SN 2012fr from the $Swift$ UVOT and the Li-Jiang 2.4-m telescope. Our observations show that SN 2012fr is a luminous SN Ia. The maximum bolometric luminosity deduced from the UV and optical light curves is $(1.82\pm 0.15) \times 10^{43}$ erg s$^{-1}$, corresponding to a synthesized nickel mass of 0.88 $\pm$ 0.08 M$_{\sun}$.

Generally, the spectral evolution of SN 2012fr show some similarities to the HV SNe Ia in the early phase because of showing detached HVFs but it becomes more similar to the 91T-like subclass in the near-maximum and later phases. In the very earlier phases, strong HVFs are present in the \ion{Ca}{2} IR triplet, \ion{Ca}{2} H\&K, and \ion{Si}{2} $\lambda$6355 lines at velocities of 22,000--25,000 km s$^{-1}$. The absorption of \ion{Si}{2} and \ion{Ca}{2} formed from the photosphere has a velocity of 12,000 km s$^{-1}$ and exhibits an unusually narrow line profile.

A comparison with other SNe Ia indicates that SN 2012fr is characterized by narrow-lined, photospheric component near the maximum light. We found that the SNe Ia similar to SN 2012fr are usually slow-decliners with $\Delta m_{15}$ (B)$\lesssim$ 1.0 mag, and they show a large overlap with the members of the shallow-silicon/SN 1991T-like subclasses in the Branch et al. and Wang et al. classification schemes. These results, together with the asymmetric high-velocity material, suggest that SN 2012fr may represent a subset of 91T-like SNe Ia viewed at different angles. A larger sample of SN 2012fr-like explosions with very early spectral observations as well as polarization measurements will help us understand the frequency of objects that show detached HVFs and their geometry (e.g., Childress et al. 2014), which will finally enable us to set stringent constraints on the nature of the progenitor system for some particular types of SNe~Ia.

\acknowledgments
We thank very much  the anonymous referee for his/her constructive suggestions which helped to improve the paper a lot. We are also extremely grateful to the staff of Li-Jiang Observatory, Yunnan Observatories of China for the observation and technological support, particularly Yu-Feng Fan, Wei-Min Yi, Chuan-Jun Wang, Yu-Xin Xin, Jiang-Guo Wang, Liang Chang and Shou-Sheng He. We acknowledge the use of public data from the $Swfit$ data archive and thanks to Bryan Irby of NASA Goddard Space Flight Center for the kind help of HEASoft installation.  Thanks goes to Yan Gao (YNAO) for the literary suggestion and modification. And we are also very grateful to Bernard Heathcote ($t\approx-13$ days), and Terry Bohlsen ($t\approx-12, -11$ days) who uploaded their earlier-phase spectra of SN 2012fr to ARAS. X. Wang is supported by the Major State Basic Research Development Program (2013CB834903), the National Natural Science Foundation of China (NSFC grants 11073013, 11178003, 11325313) and the Foundation of Tsinghua University (2011Z02170). The work of J. M. Bai is supported by the National Natural Science Foundation of China (NSFC grants 11133006, 11361140347) and the Strategic Priority Research Program ``The Emergence of Cosmological Structures" of the Chinese Academy of Sciences (grant No. XDB09000000).  B. Wang is supported by National Natural Science Foundation of China (NSFC grants 11322327 and 11103072). T. Zhang is supported by National Natural Science Foundation of China (NSFC grant 11203034). Funding for the LJ 2.4-m telescope has been provided by CAS and the People's Government of Yunnan Province.


\begin{thebibliography}{}
\bibitem[Arnett(1982)]{Arn82} Arnett, W. D. 1982, \apj, 253, 785
\bibitem[Benetti(2005)]{Ben05} Benetti, S. 2005, \apj, 623, 1011
\bibitem[Bessell(1992)]{Bess92}Bessell, M.S., 1992, in IAU Colloquium no. 136 `Stellar Photometry - Current Techniques and Future Developments',  ed.  Butler, C., $\&$ Elliott, I. p22
\bibitem[Blondin et al.(2012)]{Blondin12} Blondin,S.,   Matheson,T.,  Kirshner, R. P., et al. 2012, \aj, 143, 126
\bibitem[Branch et al.(2005)]{Branch05} Branch, D., et al. 2005, \pasp, 117, 545
\bibitem[Branch et al.(2006)]{Branch06} Branch, D., et al. 2006, \pasp, 118, 560
\bibitem[Branch et al.(2009)]{Branch09} Branch, D., Dang, L.C., $\&$ Baron, E. 2009, \pasp, 121, 238
\bibitem[Brown et al.(2009)]{Peter09} Brown, P.J., Holland, S.T., Immler, S., et al. 2009, \aj, 137, 4517
\bibitem[Brown et al.(2012)]{Peter12}Brown, P.J., Dawson, K.S.,  $\&$ Pasquale, M.  2012, \apj, 753, 22
\bibitem[Cardelli et al.(1998)]{Cardelli} Cardelli, J. A., Clayton, G. C., $\&$ Mathis, J. S. 1989, \apj, 345, 245
\bibitem[Childress et al.(2012)]{CEBT3277}Childress, M.J., Zhou, G., Tucker, B., et al. 2012, CBET, 3275, 2
\bibitem[Childress et al.(2013)]{Child12fr}Childress, M.J., Scalzo, R.A., Sim, S., et al., 2013, \apj, 770, 29 (C13)
\bibitem[Childress et al.(2014)]{ChildHVF} Childress, M.J.,Filippenko, A.V., Ganeshalingam, M., et al. 2014, \mnras, 437, 338
\bibitem[Contardo et al. (2000)]{cont00} Contardo, G., Leibundgut, B., Vacca, W. D. 2000, \aap, 359, 876
\bibitem[Cousins (1981)]{Cousins}Cousins, A. 1981, South African Astron. Obs. Circ., 6, 4
\bibitem[Dilday et al.(2012)]{Dilday} Dilday, B., Howell, D.A., Cenko, S.B., et al. 2012, Science, 337, 942
\bibitem[Filippenko et al.(1992a)]{Filip92a} Filippenko, A.V. 1992a, \apj, 384, L15
\bibitem[Filippenko et al.(1992b)]{Filip92b} Filippenko, A.V. 1992b, \aj, 104, 1543
\bibitem[Filippenko(1997)]{Filip97}Filippenko, A.V. 1997, \araa, 35, 309
\bibitem[Fisher et al.(1999)]{Fisher99}Fisher, A., Branch, D., Hatano, K., $\&$ Baron, E. 1999, MNRAS, 304, 67
\bibitem[Foley et al. (2012)]{foley12} Foley, R. J., Challis, P. J., Filippenko, A. V., et~al. 2012, \apj, 744, 38
\bibitem[Foley et al. (2013)]{foley13} Foley, R. J., Challis, P.J., Chornock, R.,  et~al. 2013, \apj, 767, 57
\bibitem[Ganeshalingam et al. (2010)]{gane10} Ganeshalingam, M., et al. 2010, \apjs, 190, 418 	
\bibitem[Gehrels et al.(2004)]{Swift04}Gehrels, N., Chincarini, G., Giommi, P. et al. 2004, \apj, 611, 1005
\bibitem[Goldhaber et al.(2001)]{Gold01}Goldhaber, G.; Groom, D. E.; Kim, A., et al. 2001, \apj,  558, 359
\bibitem[Guy et al.(2005)]{Guy05}Guy, J.; Astier, P.; Nobili, S., et al. 2005, \aap,  443, 781
\bibitem[Hamuy et al.(2003)]{Hamuy} Hamuy, M., et al. 2003, Nature, 424, 651
\bibitem[Hicken et al. (2009)]{Hick09}Hicken, M., Challis, P., Jha, S., et al. 2009, \apj, 700, 331
\bibitem[Hillebrandt $\&$ Niemeyer(2000)]{HiNi}Hillebrandt, W., $\&$ Niemeyer, J.  2000, \araa, 38, 191
\bibitem[Hsiao et al.(2007)]{Hsiao}Hsiao, E.Y., Conley, A., Howell, D.A., et al. 2007, \apj, 663, 1187
\bibitem[Johnson et al.(1966)]{Johnson}Johnson, H., Iriarte, B., Mitchell, R., $\&$ Wisniewskj, W. 1966, Commun. Lunar Planet. Lab., 4, 99
\bibitem[Kasen (2006)]{Kasen06}Kasen, D. 2006, \apj, 649, 939
\bibitem[Klotz et al.(2012)]{Klotz}Klotz, A., Normand, J., Conseil, E., Parker, S., Fabrega, J., $\&$ Maury, A. 2012, Central Bureau Electronic Telegrams, 3276
\bibitem[Li et al.(2003)]{Li02cx}Li, W., Filippenko, A., Ryan, C., et al. 2003, \pasp, 115, 453
\bibitem[Lindblad (1999)]{Lin99} Lindblad, P. O. 1999, \araa, 9, 221
\bibitem[Maund et al.(2013)]{Maund12frp}Maund, J.R., Spyromilio,  J.,  $\&$ H\"{o}flich, P.A. 2013, \mnras, 433, L20
\bibitem[Landolt (1992)]{Landolt}Landolt, A. V. 1992, AJ, 104, 340
\bibitem[Maguire et al.(2012)]{Mag2012}Maguire, K., Sullivan, M., Ellis R.S., et al. 2012, \mnras, 426, 2359
\bibitem[Maoz et al.(2014)]{Maoz14} Maoz, D., Mannucci, F., $\&$ Nelemans, G. 2014,  to appear in \araa, arXiv:1312.0628
\bibitem[Marion et al. (2013)]{marion13} Marion, G. H., et~al. 2013, \apj, 777, 40
\bibitem[Nugent et al.(1995)]{Nug95}  Nugent, P.,  Baron, E., Branch, D., et al. 1997, \apj, 485, 812
\bibitem[Pereira et al.(2013)]{Pereira11fe} Pereira, R., Thomas, R., Aldering, G., et al. 2013 \aap, 554, A27
\bibitem[Perlmutter et al.(1999)]{Perl99} Perlmutter, S.,  Aldering, G., Goldhaber, G., et al. 1999, \apj, 517, 565
\bibitem[Phillips(1992)]{Phillip92} Phillips, M. et al.1992, \aj, 103, 1632
\bibitem[Phillips(1993)]{Phillip93} Phillips, M. 1993, \apj, 413, L105
\bibitem[Phillips(1999)]{Phillip99} Phillips, M., Lira, R., Suntzeff. N.B., et al. 1999, \aj, 118, 1766
\bibitem[Poole et al.(2008)]{UVOTcali}Poole, T., Breeveld,  A., Page, M., et al. 2008, \mnras, 383, 627
\bibitem[Riess et al.(1996)]{Riess96}Riess, A., Press, W., and Kirshner, R. 1996, \apj,  473, 88
\bibitem[Riess et al.(1998)]{Riess98} Riess, A., Filippenko, A.V., Challis, P., et al. 1998, \aj, 116, 1009
\bibitem[Roming et al.(2005)]{UVOT05} Roming, P., Kennedy, T., Mason, K., et al. 2005, Space Sci. Rev., 120, 95
\bibitem[R\"{o}pke(2005)] {Ropke05} R\"{o}pke., F.K. 2009, \apj, 704, 255
\bibitem[Scalzo et al.(2010)]{Scalzo} Scalzo, R., Aldering, G., Antilogus, P., et al. 2010, \apj, 713, 1073
\bibitem[Sahu et al.(2011)]{Schu09ig} Sahu, D., Anupama, G., Anto, P., and Gurugubelli U. 2011, Proceedings of the International Astronomical Union, 7, 316
\bibitem[Schmidt et al.(1998)]{Schmidt98}Schmidt, B., Suntzeff, N., Phillips, M., et al. 1998, \apj, 507, 46
\bibitem[Schlegel et al.(1998)]{Schle98}Schlegel, D., Finkbeiner, D., and Davis, M., 1998, \apj, 500, 525
\bibitem[Silverman et al.(2011)]{Silver11} Silverman, J.,  Ganeshalingam, M.,  Li, W., et al.  2011, \mnras, 410, 585
\bibitem[Silverman et al.(2012)]{Silver12} Silverman, J.,  Foley, R.J.,  Filippenko, A.V., et al.  2012, \mnras, 425, 1789
\bibitem[Stetson (1987)]{Stetson}Stetson, P. 1987, \pasp, 99, 191
\bibitem[Stritzinger et al.(2002)]{Scor}Stritzinger, M.,  et al. 2002, \aj,  124, 210
\bibitem[Stritzinger \& Leibundgut(2005)]{56Ni05}Stritzinger, M., \& Leibundgut, B. 2005, \aap, 431, 423
\bibitem[Suntzeff et al.(1996)]{Sun96}Suntzeff, N. 1996, in Supernovae and Supernova Remnants, ed. McCray, R.,  $\&$ Wang, Z. (Cambridge:Cambridge Univ.Press), p41
\bibitem[Tully et al.(2009)]{Tully}Tully et al. 2009, AJ, 138, 323
\bibitem[Wang \& Han(2012)]{Bow12} Wang, B., $\&$ Han, Z.W. 2012, New A Rev., 56, 122
\bibitem[Wang et al. (2007)]{wanglf07} Wang, L., Baade, D., Patat, F. 2007, Science, 315, 212
\bibitem[Wang et al. (2006)]{wangxf06} Wang, X., Wang, L., Pain, R., et al. 2006, \apj, 645, 488
\bibitem[Wang et al.(2008)]{wang08}Wang, X., Li, W., Filippenko, A.V., et al. 2008, \apj, 677, 1060
\bibitem[Wang et al.(2009a)]{Wang09a} Wang, X., Filippenko, A.V., Ganeshalingam, M., et al.  2009a, \apj, 699, L139
\bibitem[Wang et al.(2009b)]{Wang09b} Wang, X., Li, W., Filippenko, A., et al.  2009b, \apj, 697, 408
\bibitem[Wang et al.(2013)]{Wang13} Wang, X., Wang, L., Filippenko, A., et al.  2013, Science, 340, 170
\bibitem[Zhang et al.(2012)]{JJzhang12} Zhang, J. J., Fan. Y.F., Chang, L., et al. 2012, Astronomical Research Technology, 9, 411
\end{thebibliography}
\end{document}